%% file: main.tex
\documentclass[letterpaper,10pt]{article}
\usepackage{fullpage}

\usepackage[dvipsnames]{xcolor}

\usepackage[bookmarks=true,pdfstartview=FitH,colorlinks,linkcolor=Cerulean,filecolor=Cerulean,citecolor=Cerulean,urlcolor=Cerulean]{hyperref}
\hypersetup{pdftitle={Bajillion: Cryptographic Clearing for Micropayments}, pdfauthor={Patrick O'Grady, Guru-Vamsi Policharla}}

\usepackage[T1]{fontenc}
\usepackage{charter}
\usepackage{amsmath}
\usepackage{amssymb}
\usepackage{amsthm}
\usepackage{amsfonts}
\usepackage{enumitem}
\usepackage{url}
\usepackage{array}
\ExplSyntaxOn
\cs_new_eq:NN \bajillion_newcounter:n \newcounter
\cs_set_protected:Npn \newcounter #1 {
  \str_case:nnT {#1} {
    {Hpclinenumber}{}
    {H@pclinenumber}{}
    {Hpcgamecounter}{}
    {Hpcrlinenumber}{}
  } {\cs_undefine:c {the#1}}
  \bajillion_newcounter:n {#1}
}
\ExplSyntaxOff
\usepackage[
lambda,
operators,
advantage,
sets,
adversary,
landau,
probability,
notions,
logic,
ff,
mm,
primitives,
events,
complexity,
asymptotics,
keys,
oracles]{cryptocode}
\ExplSyntaxOn
\cs_set_eq:NN \newcounter \bajillion_newcounter:n
\cs_undefine:N \bajillion_newcounter:n
\ExplSyntaxOff
\usepackage{mdframed}
\usepackage[capitalise]{cleveref}
\usepackage{bm}

\newtheorem{theorem}{Theorem}
\newtheorem{claim}{Claim}
\newtheorem{remark}{Remark}
\newtheorem{definition}{Definition}

\input{macros.tex}

\title{Bajillion: Cryptographic Clearing for Micropayments}
\author{Patrick O'Grady\qquad Guru-Vamsi Policharla\\[4pt] Commonware, Inc.}

\begin{document}
\date{}

\maketitle

\begin{abstract}
  Much of the web is funded by advertising, a model that erodes as humans delegate browsing to AI agents acting on their behalf.
  A natural replacement is for agents to pay for what they consume, often a fraction of a cent per request. At the scale of web traffic, this could mean tens or even hundreds of millions of payments per second.

  We present Bajillion, a \emph{clearing} protocol for scalable micropayments. A sender signs a payment request, and an untrusted operator answers with a \emph{preconfirmation}, which promises the receiver that it will be paid. Periodically, the operator closes an epoch by posting only the net effect of its payments on chain.
  Bajillion guarantees that a malicious operator can only omit payments -- it can never debit a sender beyond what it signed, nor misdirect a transfer. Posted balances are therefore finalized immediately and can be withdrawn. By design, senders never have to raise disputes, making it suitable for ephemeral AI agents.
  A receiver that is shortchanged can prove that a close omitted a preconfirmed payment, which halts the protocol and lets every user exit with its finalized balance. Requests from different senders can be processed independently, allowing the operator to parallelize up to the number of senders. We capture these guarantees in an ideal functionality for accountable clearing and prove that Bajillion realizes it.
\end{abstract}

\input{introduction.tex}

\input{preliminaries.tex}
\input{definitions.tex}
\input{construction.tex}
\input{security.tex}

\bibliographystyle{alpha}
\bibliography{abbrev3,crypto,extra,ai-extra}

\end{document}

%% file: macros.tex
\providecommand{\IfDarkModeTF}[2]{#2}
\providecolor{fgcolor}{gray}{0}
\makeatletter
\IfDarkModeTF{\colorlet{pagebg}{darkmode@anthrazitgrau}}{\colorlet{pagebg}{white}}
\makeatother
\colorlet{Shade}{gray!20!pagebg}
\newcommand{\hlt}[1]{{\setlength{\fboxsep}{2pt}\colorbox{Shade}{\ensuremath{#1}}}}
\def\cF{{\mathcal{F}}}
\def\cK{{\mathcal{K}}}
\def\cN{{\mathcal{N}}}
\def\cP{{\mathcal{P}}}
\def\cZ{{\mathcal{Z}}}

\newcommand{\bbN}{\mathbb{N}}

\newcommand{\REAL}{\mathsf{REAL}}
\newcommand{\IDEAL}{\mathsf{IDEAL}}
\newcommand{\Sim}{\mathsf{Sim}}

\newcommand{\Verify}{\mathsf{Verify}}
\newcommand{\Com}{\mathsf{Com}}

\newcommand{\cmtbody}[1]{\hspace{0.6em}{\footnotesize\sffamily\color{gray!65!fgcolor}//\,#1}}
\newcommand{\funcsep}{\par\medskip{\centering\rule{0.3\linewidth}{0.4pt}\par}}
\newcommand{\cmt}[1]{\ifmmode\text{\cmtbody{#1}}\else\cmtbody{#1}\fi}
\newlength{\eqcmtwidth}
\newsavebox{\eqcmtbox}
\newcommand{\eqcmt}[2]{%
  \begingroup
  \setbox0=\hbox{$\displaystyle #1$}%
  \setlength{\eqcmtwidth}{\dimexpr 0.5\linewidth - 0.5\wd0 - 1.2em\relax}%
  \sbox{\eqcmtbox}{\parbox[c]{\eqcmtwidth}{%
      \footnotesize\sffamily\color{gray!65!fgcolor}\raggedright
      \setbox2=\hbox{//\,}\hangindent=\wd2 \hangafter=1
      \noindent//\,#2\par}}%
  \[#1\tag*{\makebox[1pt][r]{\usebox{\eqcmtbox}}}\]%
  \endgroup}

\newcommand{\KeyGen}{\mathsf{KeyGen}}
\newcommand{\Sig}{\mathsf{Sig}}
\newcommand{\Sign}{\mathsf{Sign}}
\newcommand{\Send}{\mathsf{Send}}
\newcommand{\Close}{\mathsf{Close}}
\newcommand{\Challenge}{\mathsf{Challenge}}

\newcommand{\Fclearing}{\mathcal{F}_{\mathsf{clearing}}}
\newcommand{\Fvalidity}{\mathcal{F}_{\mathsf{validity}}}
\newcommand{\Valid}{\mathsf{Valid}}
\newcommand{\MTOpen}{\mathsf{MTOpen}}
\newcommand{\Release}{\mathsf{Release}}
\newcommand{\Init}{\mathsf{Init}}
\newcommand{\Auth}{\mathsf{Auth}}
\newcommand{\Pipe}{\mathsf{Pipe}}
\newcommand{\Carry}{\mathsf{Carry}}
\newcommand{\eid}{\mathsf{eid}}
\newcommand{\cur}{\mathsf{cur}}
\newcommand{\nxt}{\mathsf{nxt}}

\newcommand{\Bal}{\mathsf{Bal}}
\newcommand{\Rcpt}{\mathsf{Rcpt}}
\newcommand{\Dep}{\mathsf{Dep}}
\newcommand{\Wdr}{\mathsf{Wdr}}
\newcommand{\FWdr}{\mathsf{FWdr}}
\newcommand{\Req}{\mathsf{Req}}

\newcommand{\cnt}{\mathsf{cnt}}
\newcommand{\minreq}{\mathsf{min}}

\newcommand{\vroot}{\mathsf{root}}
\newcommand{\MT}{\mathsf{MT}}
\newcommand{\Root}{\mathsf{Root}}
\newcommand{\ProveMem}{\mathsf{ProveMem}}
\newcommand{\ProveNM}{\mathsf{ProveNM}}
\newcommand{\VerifyMem}{\mathsf{VerifyMem}}
\newcommand{\VerifyNM}{\mathsf{VerifyNM}}
\newcommand{\Deposit}{\mathsf{Deposit}}
\newcommand{\ForcedWithdraw}{\mathsf{ForcedWithdraw}}
\newcommand{\SoftWithdraw}{\mathsf{SoftWithdraw}}
\newcommand{\Refund}{\mathsf{Refund}}
\newcommand{\Fault}{\mathsf{Fault}}
\newcommand{\op}{\mathsf{op}}
\newcommand{\live}{\mathsf{live}}

\newcommand{\com}{\mathsf{com}}

\mdfdefinestyle{protocolbox}{
  linecolor=fgcolor,
  backgroundcolor=pagebg,
  fontcolor=fgcolor,
  frametitlefontcolor=fgcolor,
  linewidth=0.4pt,
  innerleftmargin=1em,
  innerrightmargin=1em,
  innertopmargin=0.6em,
  innerbottommargin=0.6em,
  skipabove=0.75em,
  skipbelow=0.4em,
  frametitlealignment=\centering,
  frametitlefont=\normalfont\bfseries,
  frametitleaboveskip=0.4em,
  frametitlebelowskip=0.4em,
  frametitlerule=false,
  splittopskip=1.2em,
  splitbottomskip=0.6em
}

\newenvironment{breakableprotocol}[3]{%
  \refstepcounter{figure}\label{#3}%
  \def\breakableprotocolcaption{#2}%
  \begin{mdframed}[style=protocolbox, frametitle={#1}]%
}{%
  \end{mdframed}%
  \par\noindent{\centering\figurename~\thefigure: \breakableprotocolcaption\par}%
}

\newlist{hybrids}{enumerate}{10}
\setlist[hybrids]{label=\textbf{H}$_\arabic*$\textbf{:},ref=\textbf{H}$_\arabic*$}

%% file: introduction.tex
\section{Introduction}
Much of the web is funded by advertising, and in the United States alone, internet advertising revenue reached \$294.6 billion in 2025~\cite{iab-pwc26}. As humans increasingly delegate their online presence to AI agents that read pages and take actions on their behalf, this model is quickly eroding.
Most AI platforms crawl pages far more often than they refer readers back to them~\cite{cloudflare-crawl25}, and automated traffic is growing about eight times faster than human traffic~\cite{human26}.
Sites that depend on advertising have started to push back. Amazon, which earned \$68.6 billion worldwide from advertising in 2025~\cite{amzn-10k25}, blocks dozens of AI crawlers and agents in its robots.txt~\cite{amzn-robots26}.

A natural replacement is for agents to pay for what they consume, often a fraction of a cent per request. Hyperscalers and CDNs have begun to integrate support for this natively. Cloudflare and AWS already let site owners answer an agent's request with an HTTP 402 (Payment Required) response and serve the content once the agent pays~\cite{cloudflare-ppc25,cloudflare-mg26,aws-waf26}. Cloudflare alone serves over 81 million HTTP requests per second on average~\cite{cloudflare-radar25}, so if a sizeable fraction of web requests were paid for, these payments would have to be settled at an enormous rate. This raises the question that we tackle in this work:
\begin{quote}
  \centering\emph{How do we scale to hundreds of millions of micropayments per second, each as small as \$0.000001?}
\end{quote}

Larger payments made by agents, such as booking travel or ordering groceries, can run on traditional rails by giving the agent secure access to a credit card. This may even be preferable, since such payments inherit the card networks' protections against fraud and theft. Micropayments, however, have a very different risk profile. Arbitration is not feasible for such tiny amounts, as resolving a dispute costs far more than the payment itself, and it is cheaper for a buyer to simply never return to a misbehaving seller, e.g., by maintaining a local denylist. The protections of card networks come with large fees associated with them (around 3\% plus 30 cents per transaction~\cite{stripe-pricing}) but are of little use for micropayments.

The next logical attempt would be to settle micropayments as stablecoin payments on a blockchain. Blockchains offer instant settlement without chargebacks, and only charge a flat fee for transfers of any size. In fact, Cloudflare's Monetization Gateway and AWS currently settle agent payments in stablecoins on a blockchain~\cite{cloudflare-mg26,aws-waf26} via x402~\cite{x402}.
There has also been a push to scale the base layer itself~\cite{flow20,narwhal22,CCS:SGSK22,FC:SAGL24,CCS:BCDKKL24,FC:MalStaYin24,mirbft19,NDSS:BCDKKK25,aleph19,TCC:ChaPas23,autobahn24,shoalpp25,SP:SSKN25,mahimahi25,zaptos25,monadbft25,raptr25,alpenglow25,kudzu25,hydrangea26,minimmit25,bluebottle25,cadence26,multimmit} which has made substantial progress. Recent protocols order about a million transactions per second~\cite{multimmit,multimmit-blog}, enough for every person in the world to make 10 payments a day (on average). Raising validator requirements or sharding the network are plausible ways to scale further, so it is conceivable that blockchains reach the required throughput.

However, the main obstacle is the fee paid by each transaction. A transaction on a blockchain must be disseminated to the entire network, verified, executed, and stored. Each micropayment is worth so little that this can cost more than the payment itself. Agentic micropayments therefore need an \emph{off-chain} scaling mechanism, where the net effect of micropayments made over an epoch is posted on chain instead of every payment.

We now discuss the agentic micropayment setting. The sender (buyer) is a short-lived program created to solve a given task. It pays for the resources it needs, solves the problem assigned to it, and disappears. It is unreasonable to expect the agent to come back online for checks and disputes. We assume receivers (sellers) are long-lived services capable of raising disputes. We allow an \emph{untrusted} operator to facilitate payments by issuing \emph{preconfirmations} for transfer requests, which promise the receiver that it will be paid.

We set the following goals for a clearing mechanism intended to be used for micropayments:
\begin{enumerate}
  \item \emph{Senders Never Dispute.} Senders should not be tasked with monitoring the chain or challenging the operator and must never be debited more than they authorized.
  \item \emph{Immediate Finalization.} As the operator posts data to the chain after \emph{netting} transactions over an epoch, posted balances must be immediately final, so that users can spend them in the very next epoch or withdraw them.
  \item \emph{Operator Scales with Senders.} The operator is the bottleneck in this system as every single micropayment request flows through it. We demand that the operator can be parallelized up to the number of \emph{senders}, so that transfer requests from different senders can be processed entirely independently of each other even if they involve the same receiver.
  \item \emph{Guaranteed Exit.} We do not provide \emph{liveness} guarantees against malicious operators, but users must be able to withdraw their funds and exit the clearing protocol at any time, even if the operator is unresponsive or deviates arbitrarily to prevent withdrawals.
\end{enumerate}

\paragraph{Cryptographic clearing.} Our approach is \emph{clearing}, as in traditional finance. A card merchant receives an authorization at once and is paid later, when the day's transactions are settled in a batch. The clearing house in the middle, however, is centralized and trusted. Bajillion replaces it with a single untrusted operator, which issues preconfirmations and periodically closes an epoch, and a validity oracle, which checks every close before it reaches the chain. The oracle can be realized by a committee with an honest majority.

Concretely, users join by depositing funds into a contract on the chain. To pay, a sender signs a request and sends it to the operator, which checks the sender's balance and countersigns the request as a preconfirmation. The request commits to the cumulative amount that the sender has paid to each receiver in the epoch, so the operator only needs the sender's own state to answer it and can process requests from different senders independently. The sender forwards the countersigned request to the receiver as a receipt. At the end of each epoch, the operator closes it by submitting the net transfers to the validity oracle, which checks them and posts the resulting balances on chain. Users can withdraw through the operator, or directly through the contract if the operator does not cooperate.

We guarantee that even if the operator is fully malicious:
\begin{itemize}
  \item operator misbehavior is restricted to the omission of preconfirmations:
        \begin{itemize}
          \item the operator cannot mint or double spend funds
          \item funds cannot be stolen from offline users -- any debit requires explicit authorization
          \item authorized transfers cannot be \emph{misdirected} to a different receiver
        \end{itemize}
  \item receivers can prove that they were shortchanged by the operator, which halts the protocol and holds the operator accountable
\end{itemize}

We capture these guarantees in an ideal functionality for accountable clearing, in which every preconfirmed payment is finalized or the system halts, and prove that Bajillion realizes it. Since the operator can only omit payments, a close is final as soon as it is posted, with no challenge window. If a receiver later proves that a close omitted one of its preconfirmed payments, the system halts. A halt returns finalized balances and does not restore the omitted payment, so each receiver chooses how far to rely on preconfirmations before a close.

\paragraph{Related work.} Existing off-chain scaling mechanisms are unsatisfactory for micropayments. Payment channels~\cite{spilman13,dmc15,lightning16,eltoo18} lock liquidity for each sender-receiver pair, so an agent must fund a separate channel for every service it pays, and opening each channel costs an on-chain transaction. Channel networks~\cite{lightning16,FC:MBBKM19,NDSS:MMSKM19,CCS:EggMorMaf19,USENIX:AMKM21}, virtual channels~\cite{CCS:DziFauHos18,EC:DEFHH19,SP:AMEEFR21}, and payment hubs~\cite{NDSS:HABSG17,CCS:GreMie17,SP:DEFM19,SP:TaiMorMaf21,SP:QPMSES23,CCS:GGWLXG23} route payments through intermediaries instead, but an intermediary must still lock funds in a channel to every receiver it serves.

Plasma~\cite{plasma17,plasmamvp18,plasmacash18} and the commit-chain NOCUST~\cite{EPRINT:KZFMG18} avoid per-pair liquidity, but they are secure only if every user with a balance, senders included, regularly checks the operator's commitments and challenges or exits in time, and withdrawals must wait out a challenge window.
In NOCUST-ZKP, a variant of NOCUST, the operator instead attaches a succinct proof of validity to every commitment, so users no longer need to check that commitments are valid.
Bajillion shares similarities with commit-chains, but the two differ in what a sender signs, and hence in what the operator commits to. A NOCUST checkpoint records every single payment as a separate entry, and its operator must coordinate all payments to the same receiver, so requests from different senders cannot be processed independently and a popular receiver becomes a point of contention. In Bajillion, a sender instead signs the cumulative amount it has paid to each receiver, so requests from different senders are independent and a close contains a single entry for every pair of sender and receiver, no matter how many payments were made. When senders make many payments to the same receiver, this reduces the data that the committee must check and store, and likewise the size of the witness in the variant with succinct proofs. We give a more detailed comparison against commit-chains in \cref{sec:extensions}.

Finally, recent protocols such as x402, MPP, and AP2 standardize how an agent learns a price, authorizes a payment, and presents it to a seller~\cite{x402,mpp,ap2}. However, they are agnostic to the underlying settlement layer, and Bajillion can serve as an efficient and highly scalable settlement layer for them.

%% file: preliminaries.tex
\section{Preliminaries}\label{sec:prelim}

We review the security framework and cryptographic primitives used to authenticate cumulative payments in our construction.

\paragraph{Simulation-based security.} We prove security via simulation, following the standard real-world versus ideal-world paradigm. A cryptographic scheme specifies an interactive protocol $\Pi$ among parties $P_1,\ldots,P_n$. The protocol is meant to emulate an ideal functionality $\cF$ that takes inputs from the parties and delivers outputs to them. In our setting the parties are the users, the operator $P_\op$, and the ledger $P_L$, which is never corrupted. Our functionality is \emph{reactive}: parties receive inputs over the course of the execution, and the adversary schedules its progress through dedicated commands. An \emph{environment} $\cZ$ chooses these inputs adaptively. It takes as input the security parameter $1^\secpar$ and an auxiliary input $z$. It gives each honest party its inputs one at a time, sees every output the honest parties produce, and may exchange messages with the adversary throughout. At the end, $\cZ$ outputs a bit.

\smallskip\noindent\emph{The real execution.} The protocol $\Pi$ runs in the presence of an adversary $\adv$ that corrupts a subset $M\subset[n]$ of the parties. $\adv$ takes as input $1^\secpar$. The honest parties follow $\Pi$ on the inputs $\cZ$ gives them, while $\adv$ sends messages on behalf of the parties in $M$ following an arbitrary polynomial-time strategy. The output of $\cZ$ defines the random variable $\REAL_{\Pi,\adv,\cZ}[1^\secpar,z,M]$.

\smallskip\noindent\emph{The ideal execution.} A simulator $\Sim$ controlling the parties in $M$ interacts with a trusted party running $\cF$. $\Sim$ takes as input $1^\secpar$ and has black-box access to $\adv$, written $\Sim^\adv$: it runs $\adv$ as a subroutine and sees only the messages $\adv$ sends, without rewinding $\adv$ or reading its random tape. Each honest party forwards every input it receives from $\cZ$ to $\cF$, and outputs to $\cZ$ whatever $\cF$ outputs to it. $\Sim$ sends inputs and commands to $\cF$ on behalf of the parties in $M$, receives the outputs $\cF$ delivers to them, and relays the messages between $\adv$ and $\cZ$. The output of $\cZ$ defines the random variable $\IDEAL_{\cF,\Sim^\adv,\cZ}[1^\secpar,z,M]$.

\begin{definition}[Secure emulation]\label{def:emulation}
  An $n$-party protocol $\Pi$ \emph{securely emulates} an ideal functionality $\cF$ in the presence of malicious adversaries corrupting a subset $M\subset[n]$ if there exists a PPT oracle machine $\Sim$ such that for every PPT malicious $\adv$ corrupting $M$ and every PPT environment $\cZ$,
  \[\{\REAL_{\Pi,\adv,\cZ}[1^\secpar,z,M]\}_{\secpar,z}\approx_c\{\IDEAL_{\cF,\Sim^\adv,\cZ}[1^\secpar,z,M]\}_{\secpar,z}.\]
\end{definition}

\paragraph{Digital signatures.} We use a signature scheme $\Sig=(\KeyGen,\Sign,\Verify)$ with standard EUF-CMA security: $\Sig.\KeyGen(1^\secpar)$ outputs a key pair $(\pk,\sk)$, $\Sig.\Sign(\sk,m)$ outputs a signature $\sigma$ on the message $m$, and $\Sig.\Verify(\pk,m,\sigma)$ outputs a bit.

\paragraph{Merkle trees.} We use a binary Merkle tree with a hash function $H$ to authenticate a list $\vec x=(x_1,\ldots,x_\ell)$.
\begin{itemize}[nosep,leftmargin=1.5em]
  \item $\MT.\Root(\vec x)\to\vroot$: a deterministic function that computes the root of the tree over $\vec x$.
  \item $\MT.\ProveMem(\vec x,i)\to\pi$: a deterministic function that computes an opening proof for the element at position $i$ of $\vec x$.
  \item $\MT.\VerifyMem(\vroot,x,\pi)\to\{0,1\}$: a deterministic function that verifies an opening proof and outputs 1 if the proof is valid, 0 otherwise.
\end{itemize}
When the hash $H$ is collision resistant, it is computationally infeasible for an adversary to produce a list $\vec x$, a value $x^\star$, and a proof $\pi^\star$ with $\MT.\VerifyMem(\MT.\Root(\vec x),x^\star,\pi^\star)=1$ such that $x^\star$ is not the element of $\vec x$ at the position determined by $\pi^\star$.

\paragraph{Payments maps and matrices.} Accounts $\cN$ are totally ordered. A \emph{payments map} $\cP:\cN\to\bbN$ gives the cumulative amount $\cP[B]$ paid to each account $B$. Its \emph{leaves} $\vec\cP$ are the pairs $(B,\cP[B])$ with $\cP[B]\neq0$, sorted by account, and $\Com(\cP)\coloneq\MT.\Root(\vec\cP)$ commits to them. An opening of a leaf under $\Com(\cP)$ authenticates one account's amount without needing the others. Likewise, the leaves $\vec S$ of a matrix $S:\cN\times\cN\to\bbN$ are the pairs $((A,B),S[A][B])$ with $S[A][B]\ne0$, sorted lexicographically by $(A,B)$. We write $\MT.\Root(S)$, $\MT.\ProveMem(S,i)$, and $\MT.\ProveNM(S,(A,B))$, defined below, for the operations on $\vec S$. Leaves and keys are encoded unambiguously, and the verifier compares keys in the same order as the one used to sort the leaves.

\paragraph{Non-membership.} Let the elements carry keys from a totally ordered set $\cK$, with $x_i=(k_i,\cdot)$. A list is \emph{sorted} if $k_1<k_2<\cdots<k_\ell$. A sorted list holds at most one element per key, and a key $k\in\cK$ that belongs to no element lies below $k_1$, above $k_\ell$, or strictly between the keys of two adjacent elements. We assume the root also binds the length $\ell$, so an opening proof determines the position of its element and, in particular, whether it is the first or last. Two further operations turn this into non-membership proofs:
\begin{itemize}[nosep,leftmargin=1.5em]
  \item $\MT.\ProveNM(\vec x,k)\to\bar\pi$: for a sorted $\vec x$ and a key $k$ of no element, output the neighbors of where $k$ would sit, each with its opening $\pi_j\gets\MT.\ProveMem(\vec x,j)$:
        \begin{itemize}[nosep]
          \item $\bar\pi=(x_i,\pi_i,x_{i+1},\pi_{i+1})$ if $k_i<k<k_{i+1}$
          \item $\bar\pi=(x_1,\pi_1)$ if $k<k_1$, and $\bar\pi=(x_\ell,\pi_\ell)$ if $k>k_\ell$
          \item $\bar\pi=()$ if $\vec x$ is empty
        \end{itemize}
  \item $\MT.\VerifyNM(\vroot,k,\bar\pi)\to\{0,1\}$: accept iff every opening in $\bar\pi$ verifies under $\vroot$ and one of the following holds:
        \begin{itemize}[nosep]
          \item $\bar\pi$ holds two elements at adjacent positions with keys $k_i<k<k_{i+1}$
          \item $\bar\pi$ holds one element at the first position with $k<k_1$, or one element at the last position with $k_\ell<k$
          \item $\bar\pi$ holds no element and $\vroot=\MT.\Root(())$, the root of the empty list
        \end{itemize}
\end{itemize}
When $H$ is collision resistant, $\vec x$ is sorted, and $\vroot=\MT.\Root(\vec x)$, it is computationally infeasible to produce a key $k$ of some element of $\vec x$ and a proof $\bar\pi$ with $\MT.\VerifyNM(\vroot,k,\bar\pi)=1$. The guarantee is conditional on sortedness, which the verifier cannot see from the root: whoever certifies a root must check that the list behind it is sorted.

%% file: definitions.tex
\section{Definitions}\label{sec:defs}

We model the security guarantees of an \emph{accountable} clearing mechanism via the simulation-based paradigm (\cref{def:emulation}). We begin with an overview of the ideal functionality $\Fclearing$~(\cref{fig:func}).

The mechanism runs between an operator $P_\op$, a deterministic ledger $P_L$ which cannot store secret information, and users that wish to pay each other. Any user that stores funds on the ledger can choose to participate in the clearing protocol by depositing funds through the ledger, which forwards the $\Deposit$ to $\Fclearing$. We intentionally abstract away the details of $P_L$, and make minimal assumptions about how it works. This allows us to integrate instantiations of $\Fclearing$ with \emph{any} blockchain that supports smart contracts.

The target application of such a clearing mechanism is agentic micropayments where we expect many agents (millions of them) to pay tiny amounts $\approx \mathdollar 0.001$ to a smaller set of valuable service providers (thousands of them). Of course one can implement agentic payments on any blockchain, but the fees charged by the chain will be comparable to or even larger than the payment itself. Moreover, the latency between firing off a payment and obtaining a confirmation ($\approx$ hundreds of ms) is simply too high to create a seamless browsing experience for agents. Instead, we turn to accountable clearing protocols which provide weaker \emph{preconfirmations} (see below) but can scale to massive throughputs with much lower resource requirements.

An accountable clearing system proceeds in \emph{epochs} which are consecutive, non-overlapping periods of arbitrary lengths. Each epoch is either open or finalized: closing an epoch finalizes it at once. When an epoch is open, users can freely transact with each other but rely on the operator for liveness. However, there are two important caveats:
\begin{itemize}
  \item \textbf{Preconfirmation.} While an epoch is open, the transfers are not yet finalized but the operator provides a \emph{preconfirmation} which guarantees that either the payment will be finalized or the protocol halts (see below): the operator is held accountable for every preconfirmation it issues. After a halt, users must be able to withdraw their balance recorded at the latest finalized epoch, together with any deposits that have not been processed.

        Since closing an epoch finalizes it at once, epochs can be short, and each receiver chooses how far to rely on preconfirmations. A receiver can act on preconfirmed payments up to an amount it is willing to lose if the protocol halts, and beyond that wait for the close of the epoch: once the close is posted and credits a payment, the payment is final.

        In practice, an operator can additionally set up an insurance fund or have some other reconciliation mechanism to ensure that sellers are made whole even if the protocol halts. While our definition does not formally require it, it can be extended to do so. For simplicity we restrict ourselves to the minimal guarantee that users can withdraw funds that have been finalized.
  \item \textbf{Censorship.} Our definition provides no censorship resistance: an operator can block any payment requests but we emphasize that the operator is \emph{non-custodial} -- they cannot move user funds without explicit authorization. Our definition also demands that a user can always withdraw their funds, i.e., the operator cannot censor requests to exit.
\end{itemize}
Within each epoch, the ideal functionality maintains five maps:
\begin{itemize}
  \item $\Auth_\eid[A][B]: \cN \times \cN \to \bbN$: the amount authorized by account $A$ to be paid to account $B$
  \item $\Pipe_\eid[A][B]: \cN \times \cN \to \bbN$: with a corrupt operator, the amount of an honest $A$'s \emph{pipelined} requests to pay $B$ in epoch $\eid$. These may settle in $\eid$ or $\eid+1$, but their amount is debited at most once. A pipelined request for an epoch that is still open is recorded here, not in $\Auth_\eid$
  \item $\Carry_\eid[A][B]: \cN \times \cN \to \bbN$: the portion of $\Pipe_\eid[A][B]$ that did not settle in $\eid$, and so may still settle in $\eid+1$. It is zero until $\eid$ is closed
  \item $\Rcpt_\eid[A][B]: \cN \times \cN \to \bbN$: the amount for which $B$ holds a \emph{preconfirmation} of payment from $A$
  \item $\Wdr_\eid[A]: \cN \to \bbN$: a \emph{soft request} by account $A$ to withdraw funds in epoch $\eid$, but it can be ignored by the operator
\end{itemize}
We maintain two different maps for authorizations and receipts because malicious operators or senders can withhold delivery for arbitrarily long, modeled by a $\Release$ interface. If both the sender and operator are honest, then $\Auth_\eid[A][B] = \Rcpt_\eid[A][B]$. Users can also make deposit and \emph{forced} withdrawal requests on the ledger, kept in order in two queues:
\begin{itemize}
  \item $\Req^\Dep$, a sequence of pairs in $\cN\times\bbN$: deposit requests made by users in order
  \item $\Req^\FWdr$, a sequence of pairs in $\cN\times\bbN$: withdrawal requests made by users in order
\end{itemize}
The close of an epoch must process every request queued before the epoch opened, and may process any number of later ones, in order.
Let $\Bal[A]$ denote the finalized balance of $A$ and $\cur$ the current epoch. The live balance of a user is defined as
\[
  \live(A)\coloneq\Bal[A]-\textstyle\sum_B\Auth_\cur[A][B]
\]
which is used by honest operators to decide whether a send request should be approved. Note that the definition does not \emph{demand} deposits and incoming transfers be spendable in the same epoch, as they make scaling much more challenging. Consider a very popular receiver $B$ who is also sending funds. In order to scale, an operator would want to massively parallelize the processing of incoming requests -- potentially up to the number of senders $\{A_i\}_i$ itself. But if incoming deposits and transfers must be immediately usable, an operator must synchronize across all of the threads before approving a transfer request by $B$.

Although the definition does not demand it, as discussed in~\cref{sec:construction}, our protocol can be easily adapted to support this feature.

\paragraph{Pipelined requests.} A sender that does not mind whether a request settles in epoch $\eid$ or $\eid+1$ can mark it pipelined. If the epoch closes before the operator answers, the request can still settle in the next epoch without being resubmitted. An honest operator approves it in whichever of the two epochs is current. A corrupt operator may settle it in either epoch.

\paragraph{Close and Fault.} Operators close an epoch via the $\Close$ interface, choosing how many queued requests the close processes. The close finalizes the epoch: the finalized balances $\Bal$ are updated and withdrawals are paid out at once.

If the operator is honest, every authorized transfer settles and each user is paid out its forced and soft withdrawals, up to its balance:
\begin{gather*}
  b[A]\coloneq\Bal[A]+\textstyle\sum_B\Auth_\cur[B][A]-\sum_B\Auth_\cur[A][B]+d_A,\\
  \Bal[A]\gets b[A]-\min(b[A],\,w_A+\Wdr_\cur[A])
\end{gather*}
where $d_A$ and $w_A$ are the sums of $A$'s deposits and forced withdrawals the epoch processes.

If the operator is corrupt, they choose a map $S[A][B]:\cN \times \cN \to \bbN$ of funds that will move from $A \to B$ and the ideal functionality enforces that:
\begin{itemize}
  \item every honest sender $A$ is debited at most what they authorized for the epoch being closed: its plain requests for this epoch, its pipelined requests for the previous epoch that did not settle there, and its pipelined requests for this epoch. The portion of the latter that does not settle carries over to the next epoch
  \item every honest receiver $B$ is credited what they were promised, $S[A][B] \geq \Rcpt_\cur[A][B]$, or the protocol halts right after the close
  \item every request queued before the epoch opened is processed, processed deposits are credited, and each processed forced withdrawal is paid in full or empties the balance
\end{itemize}
A corrupt operator can ignore soft withdrawal requests entirely and could deviate arbitrarily for the debits and credits of corrupt parties. A close that shortchanges an honest receiver still goes through: these checks ensure it cannot debit an honest user beyond what they authorized, so undoing a close would not make the receiver whole.

If the operator is corrupt, the adversary can halt the protocol at any point. The ledger can also halt it, e.g., when the operator stops responding. On a halt, the open epoch is discarded and users are paid out the latest finalized balances and any deposits that have not been processed.

\begin{breakableprotocol}{Ideal functionality $\Fclearing$}{The clearing functionality.}{fig:func}
  \noindent\textbf{State.} The finalized balances are stored in $\Bal\gets\Init$, initialized to the starting balances at the start of the protocol. Denote the operator's current epoch by $\cur\gets1$. Epochs $1,\ldots,\cur-1$ are finalized. Maintain a queue $\Req^\Dep\gets()$ of deposit requests and a queue $\Req^\FWdr\gets()$ of forced withdrawal requests, each request a pair $(A,x)$, the number $\minreq_\Dep,\minreq_\FWdr\gets0$ of requests that the close of $\cur$ must process, and, for every epoch $\eid$, maps $\Auth_\eid,\Pipe_\eid,\Carry_\eid,\Rcpt_\eid:\cN\times\cN\to\bbN$ and $\Wdr_\eid:\cN\to\bbN$ initialized to zero.

  \smallskip\noindent\textbf{Live balance.} The \emph{live balance} of $A$ in the current epoch is defined as
  \eqcmt{\live(A)\coloneq\Bal[A]-\textstyle\sum_B\Auth_\cur[A][B]}{live balances exclude deposits and incoming transfers in $\cur$}
  All inputs from honest parties, including $P_L$, are leaked to the adversary. If any require fails, output $\bot$ to the caller and leave the state unchanged. After a halt, any interaction returns $\bot$.

  \medskip\noindent$\bm{\Send}$. On input $(\Send,A,X,\eid,\mathsf{pipe})$, a request to pay $X=\{(B_i,x_i)\}_{i\in[m]}$ in epoch $\eid$, or in epoch $\eid$ or $\eid+1$ if $\mathsf{pipe}=\mathsf{true}$, from user $A$, require $m\ge1$, $B_i\in\cN$, $x_i>0$ for every $i$, the $B_i$ pairwise distinct, and $\mathsf{pipe}\in\{\mathsf{true},\mathsf{false}\}$. Let $E=\{\eid\}$ if $\mathsf{pipe}=\mathsf{false}$ and $E=\{\eid,\eid+1\}$ if $\mathsf{pipe}=\mathsf{true}$:
  \begin{itemize}[leftmargin=1.5em]
    \item $P_\op$ honest:
          \begin{itemize}[nosep]
            \item if $A$ is honest and $\Auth_e[A]\neq0$ for some $e>\eid$, nothing happens \cmt{an honest sender never goes back to an epoch before one it has already paid in}
            \item if $\cur\notin E$ or $\sum_i x_i>\live(A)$, nothing happens
            \item else update $\{\Auth_\cur[A][B_i]\gets\Auth_\cur[A][B_i]+x_i\}_{i\in[m]}$
                  \begin{itemize}[nosep,leftmargin=1em]
                    \item if $A$ is honest, set $\Rcpt_\cur[A][B_i]\gets\Auth_\cur[A][B_i]$ and output $(\mathsf{Received},\cur,A,\Auth_\cur[A][B_i])$ to every $B_i$
                          \cmt{receivers are notified immediately if senders are honest. A corrupt sender may withhold receipts, but the amount is authorized}
                  \end{itemize}
          \end{itemize}
    \item $A$ honest, $P_\op$ corrupt:
          \begin{itemize}[nosep]
            \item if $\mathsf{pipe}=\mathsf{false}$ and $\eid\geq\cur$, update $\{\Auth_\eid[A][B_i]\gets\Auth_\eid[A][B_i]+x_i\}_{i\in[m]}$

                  \cmt{$P_\op$ cannot process the transfer in a closed epoch}
            \item if $\mathsf{pipe}=\mathsf{true}$ and $\eid\geq\cur$, update $\{\Pipe_\eid[A][B_i]\gets\Pipe_\eid[A][B_i]+x_i\}_{i\in[m]}$
                  \cmt{a pipelined request may settle in $\eid$ or $\eid+1$}
            \item if $\mathsf{pipe}=\mathsf{true}$ and $\eid=\cur-1$, update $\{\Auth_\cur[A][B_i]\gets\Auth_\cur[A][B_i]+x_i\}_{i\in[m]}$
                  \cmt{$\eid$ is closed, so the request can only settle in $\cur$}
          \end{itemize}
    \item $A$ and $P_\op$ corrupt:
          \begin{itemize}[nosep]
            \item nothing happens
                  \cmt{together they can produce receipts for any amount}
          \end{itemize}
  \end{itemize}
  \cmt{Unless both $A$ and $P_\op$ are honest, notification of incoming transfers to the receiver can be delayed arbitrarily, by the sender withholding receipts or the operator refusing to respond to a request. This is modeled by Release.}

  \medskip\noindent$\bm{\Release}$. On $(\Release,A,B,\eid,y)$ from the adversary with $\eid\ge\cur$:
  \begin{itemize}[nosep,leftmargin=1.5em]
    \item unless $A$ and $P_\op$ are both corrupt, require $y\le\Auth_\eid[A][B]+\Pipe_{\eid-1}[A][B]+\Pipe_\eid[A][B]$
          \cmt{a state $A$ signs for $\eid$ may include its plain and pipelined requests for $\eid$, and its pipelined requests for $\eid-1$}
    \item if $y>\Rcpt_\eid[A][B]$, set $\Rcpt_\eid[A][B]\gets y$ and output $(\mathsf{Received},\eid,A,y)$ to $B$
  \end{itemize}
  \funcsep

  \medskip\noindent$\bm{\Deposit}$. On input $(\Deposit,A,x)$ from $P_L$:
  \begin{itemize}[nosep,leftmargin=1.5em]
    \item require $x>0$, append $(A,x)$ to $\Req^\Dep$
  \end{itemize}

  \medskip\noindent$\bm{\ForcedWithdraw}$. On input $(\ForcedWithdraw,A,w)$ from $P_L$, a request to withdraw up to $w$:
  \begin{itemize}[nosep,leftmargin=1.5em]
    \item require $w>0$, append $(A,w)$ to $\Req^\FWdr$
  \end{itemize}

  \medskip\noindent$\bm{\SoftWithdraw}$. On input $(\SoftWithdraw,A,w,\eid)$ from user $A$, a request to withdraw up to $w$ in epoch $\eid$, require $w>0$:
  \begin{itemize}[nosep,leftmargin=1.5em]
    \item if $P_\op$ honest and $\eid=\cur$, update $\Wdr_\cur[A]\gets\Wdr_\cur[A]+w$
    \item if $A$ honest, $P_\op$ corrupt and $\eid\ge\cur$, update $\Wdr_\eid[A]\gets\Wdr_\eid[A]+w$

          \cmt{$P_\op$ may pay out less, or nothing, in $\Close$}
    \item if $A$ and $P_\op$ corrupt, nothing happens
  \end{itemize}
  \funcsep

  \medskip\noindent$\bm{\Close}$. On input $(\Close,\cnt_\Dep,\cnt_\FWdr,\ldots)$ from $P_\op$, closing epoch $\cur$ and processing $\cnt_\Dep$ deposits and $\cnt_\FWdr$ forced withdrawals:
  \begin{itemize}[leftmargin=1.5em]
    \item require $\minreq_\Dep\le \cnt_\Dep\le|\Req^\Dep|$ and $\minreq_\FWdr\le \cnt_\FWdr\le|\Req^\FWdr|$
    \item remove the first $\cnt_\Dep$ requests from $\Req^\Dep$, call them $\Dep$, and the first $\cnt_\FWdr$ requests from $\Req^\FWdr$, call them $\FWdr$, and for every $A$ let $d_A\coloneq\sum_{(A,x)\in\Dep}x$ and $w_A\coloneq\sum_{(A,x)\in\FWdr}x$
    \item $P_\op$ honest: the input is $(\Close,\cnt_\Dep,\cnt_\FWdr)$ and let $S\gets\Auth_\cur$. For every $A$
          \begin{itemize}[nosep]
            \item let the balance before payouts be $b[A]\coloneq\Bal[A]+\sum_B S[B][A]-\sum_B S[A][B]+d_A$

                  \cmt{starting balance + incoming payments - outgoing payments + processed deposits}
            \item let $p[A]\coloneq\min(b[A],w_A+\Wdr_\cur[A])$
                  \cmt{payout up to requested amount or entire balance}
          \end{itemize}
    \item $P_\op$ corrupt: the input is $(\Close,\cnt_\Dep,\cnt_\FWdr,S,\Bal')$, with $S:\cN\times\cN\to\bbN$ and $\Bal':\cN\to\bbN$
          \begin{itemize}[nosep]
            \item require $S[A][B]\le\Auth_\cur[A][B]+\Carry_{\cur-1}[A][B]+\Pipe_\cur[A][B]$ for every honest $A$ and every $B$ \cmt{cannot debit more than authorized}
            \item for every $A$:
                  \begin{itemize}[nosep]
                    \item require the balance before payouts to be non-negative
                          \[
                            b[A]\coloneq\Bal[A] + \sum_B S[B][A] - \sum_B S[A][B] + d_A \geq 0
                          \]
                          \cmt{starting balance + incoming payments - outgoing payments + processed deposits}
                    \item require $\Bal'[A]\le b[A]$ and let $p[A]\coloneq b[A]-\Bal'[A]$ \cmt{the difference between $b[A]$ and final balance $\Bal'[A]$ is paid out to $A$}
                    \item require $p[A]\ge\min(w_A,b[A])$ \cmt{forced withdrawals are paid in full, or the balance is emptied}
                    \item if $A$ is honest, require $p[A] \leq w_A + \Wdr_\cur[A]$ \cmt{payout should not exceed soft + forced withdrawals}
                  \end{itemize}
          \end{itemize}
  \end{itemize}
  \begin{itemize}[nosep,leftmargin=1.5em]
    \item set $\Bal[A]\gets b[A]-p[A]$ for every $A$ \cmt{the close finalizes the epoch at once}
    \item output $(\mathsf{Paid},\cur,p[A])$ to every $A$ with $p[A]>0$ and $(\mathsf{Finalized},\cur,\Bal)$ to every party
    \item for every honest $A$ and every $B$, set \[\Carry_\cur[A][B]\gets\Pipe_\cur[A][B]-\max(0,S[A][B]-\Auth_\cur[A][B]-\Carry_{\cur-1}[A][B])\]
          \cmt{the part of $\Pipe_\cur$ not settled in $\cur$, using the budgets that expire now first, may still settle in $\cur+1$}
    \item set $\cur\gets\cur+1$, $\minreq_\Dep\gets|\Req^\Dep|$, and $\minreq_\FWdr\gets|\Req^\FWdr|$
    \item if $S[A][B]<\Rcpt_{\cur-1}[A][B]$ for some $A$ and some honest $B$, halt as in $\Fault$
          \cmt{a close that credits an honest receiver less than promised goes through, but halts the protocol}
  \end{itemize}

  \medskip\noindent$\bm{\Fault}$. On input $\Fault$ from $P_L$, or on $\Fault$ from the adversary with $P_\op$ corrupt, \emph{halt}:
  \begin{itemize}[nosep,leftmargin=1.5em]
    \item output $(\mathsf{Halted},\cur-1)$ to every party
          \cmt{the open epoch is discarded}
    \item let $d_A$ be the sum of $A$'s deposits in $\Req^\Dep$, the deposits no close has processed, and output $(\Refund,\Bal[A]+d_A)$ to every $A$ with $\Bal[A]+d_A>0$
  \end{itemize}
\end{breakableprotocol}

%% file: construction.tex
\section{Construction}
\label{sec:construction}

In this section we construct $\mathsf{Bajillion}$, a realization of $\Fclearing$ in the $\Fvalidity$-hybrid model (\cref{fig:validity}).

\paragraph{Senders.} Each sender $A$ maintains, for each epoch $\eid$, a payments map $\cP_{A,\eid}:\cN \to \bbN$, initialized to zero. $\cP_{A,\eid}[B]$ is the total amount to $B$ that the operator has countersigned for $\eid$. To pay $x_i$ to each receiver $B_i$ in a batch $X=\{(B_i,x_i)\}_{i\in[m]}$, the sender computes an updated payments map $\cP'_\eid$ with
\[
  \left\{\cP'_\eid[B_i] \gets \cP_{A,\eid}[B_i] + x_i \right\}_{i\in[m]}
\]
where the amounts of receivers not in the batch are copied over. The sender then signs the updated state:
\begin{gather*}
  \st_{A,\eid} = \left(\eid, A, \Com(\cP'_\eid),\Com(\cP_{A,\eid-1})\right) \\
  \sigma_{A,\eid}\gets\Sig.\Sign(\sk_A,\st_{A,\eid})
\end{gather*}
where $\st_{A,\eid}$ also includes a commitment to the previous epoch's payments map $\cP_{A,\eid-1}$.\footnote{Looking ahead this will enable \emph{pipelined} send requests ensuring that transfers are not disrupted at the boundaries of epochs.}
$A$ sends $(\st_{A,\eid},\sigma_{A,\eid})$ and $X$ to the operator. Once $A$ receives the operator's countersignature and checks it, $A$ sets $\cP_{A,\eid}\gets\cP'_\eid$ and sends each $B_i$ a receipt.

\paragraph{Operator.} For every sender $A$, the operator keeps the latest payments maps it has countersigned for $A$ in the current and the previous epoch, $\Auth_\cur[A]$ and $\Auth_{\cur-1}[A]$, together with the latest countersigned state
\[
  \bar\st_A=(\st_A,\sigma_A,\sigma_\op), \qquad \st_A=\left(\cur, A, \Com(\Auth_\cur[A]),\Com(\Auth_{\cur-1}[A])\right),
\]
where $\sigma_A$ is $A$'s signature and $\sigma_\op$ the operator's countersignature. It also maintains a live balance
\[
  \live(A)=\Bal[A]-\sum_B\Auth_\cur[A][B]
\]
which determines the spending limits of a certain account.
A request from $A$ carries a signed state $(\st_A,\sigma_A)$ and a batch $X=\{(B_i,x_i)\}_{i\in[m]}$. The operator parses $\st_A = (\eid, A, \com_\eid, \com_{\eid-1})$ and checks that:
\begin{itemize}
  \item $\eid = \cur$ and the state names $A$
  \item $\Sig.\Verify(\pk_A,\st_A,\sigma_A) = 1$
  \item $m\ge1$, $B_i\in\cN$ and $x_i>0$ for every $i$, the $B_i$ pairwise distinct, and $\sum_{i\in[m]}x_i\le\live(A)$
  \item $\com_{\eid-1} = \Com(\Auth_{\cur-1}[A])$
  \item $\com_\eid = \Com(\cP')$, where $\cP'$ is $\Auth_\cur[A]$ with $x_i$ added to the entry of every $B_i$
\end{itemize}
If a check fails, the operator responds with $\bot$. Otherwise it sets $\Auth_\cur[A]\gets\cP'$, decreases $\live(A)$ by $\sum_{i\in[m]}x_i$, countersigns $\sigma_\op\gets\Sig.\Sign(\sk_\op,\st_A)$, stores $\bar\st_A=(\st_A,\sigma_A,\sigma_\op)$, and returns $(\cur,\sigma_\op)$ to $A$.

We now describe the mechanism by which operators close epochs to finalize balances and advance to the next epoch.

\paragraph{Close.} To close the current epoch $\eid=\cur$, the operator sends the settlement matrix $S \gets \Auth_\eid$, together with the last countersigned state $\bar\st_A$ of every sender $A$ with $S[A]\ne0$,
\[(\Close,\eid,S,\{\bar\st_A\}_A)\]
to $\Fvalidity$. $\Fvalidity$ then checks the close and posts the updated balances to the ledger:
\begin{itemize}[nosep,leftmargin=1.5em]
  \item for every $A$ with $S[A]\ne0$, parse $\bar\st_A=(\st_A,\sigma_A,\sigma'_A)$, and require that
        \[\Sig.\Verify(\pk_A,\st_A,\sigma_A)=1, \qquad \st_A=(\eid,A,\Com(S[A]),\Com(S_{\eid-1}[A]))\]
        where $S_{\eid-1}$ is the settlement matrix of the previous close, which $\Fvalidity$ stores
  \item compute the updated balances \[b[A]\gets\Bal[A]-\sum_B S[A][B]+\sum_B S[B][A]\] and require $b[A]\ge0$ for every $A$
\end{itemize}
The full close also processes deposits and withdrawals, which we describe below. We omit them here to highlight the core settlement protocol.

Zooming out, $\Fvalidity$ provides strong guarantees on the close it accepts. Specifically, a malicious operator cannot divert funds to the wrong address nor can they debit beyond the authorized amounts. Every transfer in an accepted close was signed by its sender. The operator can still \emph{omit} transactions it provided preconfirmations for, but a receiver can prove such an omission. As a result, we can immediately finalize user balances allowing them to withdraw without a \emph{challenge period} commonly seen in optimistic protocols.

\paragraph{Pipelined sends.} We are now in a position to discuss the nuances surrounding retries. Observe that a sender may initiate a request for epoch $\eid$ and it may fail because:
\begin{itemize}
  \item an honest operator refuses because the epoch has advanced
  \item a corrupt operator refuses for an arbitrary reason
\end{itemize}
The natural recourse here is to retry in the next epoch but the sender must ensure that their request has \emph{not} been included in $\eid$'s close as otherwise a corrupt operator can include both requests and double charge the sender. In practice, this may lead to the sender waiting for a few minutes before the close is posted. To avoid this wait, we offer a \emph{pipelined} mode for send requests, where a request can be included in $\eid$ or $\eid+1$ but not in both. The idea is to sign two states for the same batch, of which at most one can pass $\Fvalidity$'s checks:
\begin{itemize}
  \item $\st_{A,\eid} = \left(\eid, A, \Com(\cP'_\eid),\Com(\cP_{A,\eid-1})\right)$
  \item $\st_{A,\eid+1} = \left(\eid+1, A, \Com(\cP'_{\eid+1}),\Com(\cP_{A,\eid})\right)$
\end{itemize}
where $\cP'_{\eid+1}$ holds only the batch, as $A$ has paid nothing in $\eid+1$ yet.
Observe that the state includes a commitment to the previous epoch's payments map. If the operator uses $\st_{A,\eid}$ in $\eid$'s close, then they cannot use $\st_{A,\eid+1}$ in $\eid+1$'s close because $\Com(\cP_{A,\eid}) \neq \Com(\cP'_\eid)$. Thus, the operator cannot include both requests in their close, which protects the sender from being charged twice.

\paragraph{Deposits and withdrawals.} Deposits and forced withdrawals go through the ledger. A user who deposits $x$ pays it to the ledger, which appends $(A,x)$ to the queue $\Req^\Dep$. A user who wants to withdraw $w$ without the operator's help appends $(A,w)$ to the queue $\Req^\FWdr$. Each close processes a prefix of both queues. It must include every request queued before the epoch opened, and may include later ones, so the operator can skip a queued request in at most one close. The close credits the processed deposits and pays out the processed withdrawals:
\[
  b[A]\gets\Bal[A]+d_A-\sum_B S[A][B]+\sum_B S[B][A], \qquad p[A]\gets\min(b[A],\,w_A+W[A]), \qquad \Bal[A]\gets b[A]-p[A],
\]
where $d_A$ and $w_A$ are $A$'s processed deposits and forced withdrawals, $W[A]$ is the total of $A$'s soft withdrawals, described below, and $p[A]$ is the payout. A deposit becomes spendable only once a close processes it, as $\live(A)$ starts from $\Bal[A]$.

Users can also withdraw through a \emph{soft} withdrawal, which needs no ledger transaction. The user $A$ picks a fresh tag $\tau$, signs the request $(\SoftWithdraw,\eid,A,w,\tau)$, and sends it to the operator. The operator collects these requests in a set $W$ and passes it to $\Fvalidity$ with the close. $\Fvalidity$ checks every signature and that no request appears twice, and $W[A]$ is the sum of $A$'s amounts in $W$. A corrupt operator may ignore a soft withdrawal, in which case $A$ falls back to a forced withdrawal. If the ledger halts, every user gets back its finalized balance and its deposits that no close has processed.

\begin{breakableprotocol}{$\mathsf{Bajillion}$}{Bajillion, in the $\Fvalidity$-hybrid model.}{fig:matrix}
  \noindent \textbf{Setup.} Every party registers a signature key with the ledger, and Merkle trees use a hash function $H$. The ledger holds the deposited funds, the finalized balances $\Bal\gets\Init$, request queues $\Req^\Dep,\Req^\FWdr\gets()$, the number of deposit and forced withdrawal requests $\minreq_\Dep,\minreq_\FWdr\gets0$ that the close for $\nxt$ must process, and $\nxt\gets1$, the epoch whose close it expects next. Epochs $1,\ldots,\nxt-1$ are finalized. The operator starts with $\cur\gets1$ and $\live\gets\Init$.

  \medskip\noindent
  $\bm{\Send}(A,X,\eid,\mathsf{pipe})$, $X=\{(B_i,x_i)\}_{i\in[m]}$, $m\ge1$, $B_i\in\cN$ pairwise distinct, $x_i>0$, $\mathsf{pipe}\in\{\mathsf{true},\mathsf{false}\}$:
  \begin{itemize}[nosep,leftmargin=1.5em]
    \item \textbf{sender:} maintains a payments map $\cP_{A,e}: \cN \to \bbN$ for each epoch $e$, the sum of its requests that the operator countersigned for $e$ or that it found in the close of $e$, initialized to zero, and at most one \emph{pending} request, ignoring $\Send$ inputs while one is pending
          \begin{itemize}
            \item if $\cP_{A,e}\neq0$ for some $e>\eid$, ignore the request
            \item let $\cP'_\eid = \cP_{A,\eid}$
            \item update $\{\cP'_\eid[B_i] \gets \cP'_\eid[B_i] + x_i\}_{i \in [m]}$
            \item let $\st_{A,\eid} = \left(\eid, A, \Com(\cP'_\eid),\Com(\cP_{A,\eid-1})\right)$
            \item sign it as $\sigma_{A,\eid}\gets\Sig.\Sign(\sk_A,\st_{A,\eid})$
            \item if $\mathsf{pipe}=\mathsf{true}$, additionally:
                  \begin{itemize}
                    \item let $\cP'_{\eid+1} = 0$
                    \item update $\{\cP'_{\eid+1}[B_i] \gets \cP'_{\eid+1}[B_i] + x_i\}_{i \in [m]}$
                    \item let $\st_{A,\eid+1} = \left(\eid+1, A, \Com(\cP'_{\eid+1}),\Com(\cP_{A,\eid})\right)$
                    \item sign it as $\sigma_{A,\eid+1}\gets\Sig.\Sign(\sk_A,\st_{A,\eid+1})$
                  \end{itemize}
            \item send the signed states and $\{(B_i,x_i)\}_{i \in [m]}$ to the operator. The request is marked pending until answered, and dropped once it has received a $\bot$ response for each of its states
          \end{itemize}
    \item \textbf{operator:} maintains, for each epoch $\eid$, a map $\Auth_\eid:\cN\times\cN\to\bbN$, where $\Auth_\eid[A]$ is the last payments map of $A$ it countersigned for $\eid$, initialized to zero (with $\Auth_0=0$). It also maintains the live balance $\live:\cN\to\bbN$, where $\live(A)=\Bal[A]-\sum_B\Auth_\cur[A][B]$. It handles each signed state $(\st_A,\sigma_A)$ in a request from $A$ with batch $X$ on its own. If a require fails, it returns $\bot$ for that state to $A$ and leaves its own state unchanged. Require $\st_A=(\cur,A,\com_A,\com'_A)$
          \begin{itemize}
            \item require $\Sig.\Verify(\pk_A,\st_A,\sigma_A)=1$
            \item require $m\ge1$, $B_i\in\cN$ and $x_i>0$ for every $i$, the $B_i$ pairwise distinct, and $\sum_{i\in[m]}x_i\le\live(A)$
            \item require $\com'_A=\Com(\Auth_{\cur-1}[A])$ \cmt{$A$ agrees on what it paid in $\cur-1$}
            \item let $\cP'=\Auth_\cur[A]$, update $\{\cP'[B_i]\gets\cP'[B_i]+x_i\}_{i\in[m]}$, and require $\com_A=\Com(\cP')$
            \item set $\Auth_\cur[A]\gets\cP'$ and $\live(A)\gets\live(A)-\sum_{i\in[m]}x_i$
            \item sign $\sigma_\op\gets\Sig.\Sign(\sk_\op,\st_A)$, store $\bar\st_A\coloneq(\st_A,\sigma_A,\sigma_\op)$, replacing the previous one, and return $(\cur,\sigma_\op)$ to $A$
          \end{itemize}
    \item \textbf{sender:} on $(e,\sigma_\op)$ for the pending request, with epoch $\eid$ and flag $\mathsf{pipe}$, check that the countersignature verifies $\Sig.\Verify(\pk_\op,\st_{A,e},\sigma_\op)=1$, and that $e=\eid$ or that $\mathsf{pipe}=\mathsf{true}$ and $e=\eid+1$. If a check fails, treat the response as $\bot$. Otherwise mark the request as answered, so it is no longer pending, and:
          \begin{itemize}
            \item set $\cP_{A,e}\gets\cP'_e$ and let $\bar\st_{A,e}\coloneq(\st_{A,e},\sigma_{A,e},\sigma_\op)$
            \item for $i\in[m]$, let $\ell_i=(B_i,\cP'_e[B_i])$ and $j_i$ its position in $\vec\cP'_e$, and send $B_i$ the \emph{receipt} $(\bar\st_{A,e},\ell_i,\pi_i)$ with $\pi_i\gets\MT.\ProveMem(\vec\cP'_e,j_i)$
          \end{itemize}
    \item \textbf{receiver:} maintains $\Rcpt_\eid:\cN\to\bbN$ for each epoch, the highest amount received from each sender, initialized to zero, together with the receipt that set it. On a receipt $(\bar\st_A,\ell,\pi)$ with $\st_A=(\eid,A,\com_A,\com'_A)$ and $\ell=(B,y)$, $B$ its own account
          \begin{itemize}
            \item require $\eid\ge\nxt$ (no close for $\eid$ has been posted), the ledger has not halted, and $y>\Rcpt_\eid[A]$
            \item require $\Sig.\Verify(\pk_A,\st_A,\sigma_A)=\Sig.\Verify(\pk_\op,\st_A,\sigma_\op)=\MT.\VerifyMem(\com_A,\ell,\pi)=1$
            \item set $\Rcpt_\eid[A]\gets y$, store the receipt and output $(\mathsf{Received},\eid,A,y)$
          \end{itemize}
  \end{itemize}

  \medskip\noindent
  $\bm{\Deposit}(A,x)$ and $\bm{\ForcedWithdraw}(A,w)$, $x,w>0$, on input to the ledger:
  \begin{itemize}[nosep,leftmargin=1.5em]
    \item \textbf{ledger:} append $(A,x)$ to $\Req^\Dep$, or $(A,w)$ to $\Req^\FWdr$
  \end{itemize}

  \medskip\noindent
  $\bm{\SoftWithdraw}(A,w,\eid)$, $w>0$:
  \begin{itemize}[nosep,leftmargin=1.5em]
    \item \textbf{user:} pick a fresh tag $\tau$, sign $q=(\SoftWithdraw,\eid,A,w,\tau)$ as $\sigma\gets\Sig.\Sign(\sk_A,q)$, and send $(q,\sigma)$ to the operator
    \item \textbf{operator:} maintains a set $W_\eid$ of soft withdrawal requests for each epoch, initially empty. On $(q,\sigma)$ with $q=(\SoftWithdraw,\eid,A,w,\tau)$, require $\eid=\cur$, $w>0$, and $\Sig.\Verify(\pk_A,q,\sigma)=1$. If $(q, \cdot) \notin W_\cur$, add $(q,\sigma)$ to $W_\cur$
  \end{itemize}

  \medskip\noindent
  $\bm{\Close}(\cnt_\Dep,\cnt_\FWdr)$, on input to the operator:
  \begin{itemize}[nosep,leftmargin=1.5em]
    \item \textbf{operator:} require $\nxt=\cur$, $\minreq_\Dep\le \cnt_\Dep\le|\Req^\Dep|$, and $\minreq_\FWdr\le \cnt_\FWdr\le|\Req^\FWdr|$, let $\eid\gets\cur$, and let $S\gets\Auth_\eid$ and $W\gets W_\eid$
          \begin{itemize}
            \item sign $\sigma_\op\gets\Sig.\Sign(\sk_\op,m)$ with $m=(\Close,\eid,\MT.\Root(S),W,\cnt_\Dep,\cnt_\FWdr)$, and send $\Fvalidity$ the \emph{close} $(\Close,\eid,S,\{\bar\st_A\}_A,W,\cnt_\Dep,\cnt_\FWdr,\sigma_\op)$
            \item compute $b'$ as $\Fvalidity$ does (\cref{fig:validity}), set $\live\gets b'$ and $\cur\gets\eid+1$, and announce the new epoch to every user
          \end{itemize}
    \item \textbf{user:} once the close for an epoch $e$ is posted, if $A$ has a pending request:
          \begin{itemize}
            \item if the request may settle in $e$, pick any receiver $B$ in its batch and obtain $\bar\pi$ from $\Fvalidity$ on $(\MTOpen,e,(A,B))$. If $\bar\pi$ opens $\cP'_e[B]$, treat the request as included in the close: mark it as answered and set $\cP_{A,e}\gets\cP'_e$
            \item if the request was not found in the close, drop it if every epoch it may settle in is at most $e$
          \end{itemize}
    \item \textbf{ledger:} maintains a root $\vroot_\eid$ of the matrix each posted close settles. The close for $\eid$ is \emph{posted}, which finalizes $\eid$, on $(\Valid,\eid,\vroot,\cnt_\Dep,\cnt_\FWdr,b',p)$ from $\Fvalidity$, and the ledger:
          \begin{itemize}
            \item set $\vroot_\eid\gets\vroot$, remove the first $\cnt_\Dep$ requests from $\Req^\Dep$ and the first $\cnt_\FWdr$ requests from $\Req^\FWdr$, and set $\minreq_\Dep\gets|\Req^\Dep|$, $\minreq_\FWdr\gets|\Req^\FWdr|$, and $\nxt\gets\eid+1$
            \item pay every $A$ with $p[A]>0$ its payout with output $(\mathsf{Paid},\eid,p[A])$, set $\Bal\gets b'$, and output $(\mathsf{Finalized},\eid,\Bal)$ to every party
          \end{itemize}
  \end{itemize}

  \medskip\noindent
  $\bm{\Challenge}$, against a posted epoch $\eid$:
  \begin{itemize}[nosep,leftmargin=1.5em]
    \item \textbf{receiver:} when the close for $\eid$ is posted, for every $A$ with $\Rcpt_\eid[A]>0$ obtain $\bar\pi$ from $\Fvalidity$ on $(\MTOpen,\eid,(A,B))$. If it opens an amount below $\Rcpt_\eid[A]$, or proves $(A,B)$ absent, post the stored receipt from $A$ together with $\bar\pi$
    \item \textbf{ledger:} on $(\bar\st_A,\ell,\pi,\bar\pi)$ with $\st_A=(\eid,A,\com_A,\com'_A)$, $\eid<\nxt$, and $\ell=(B,y)$: require $\Sig.\Verify(\pk_A,\st_A,\sigma_A)=\Sig.\Verify(\pk_\op,\st_A,\sigma_\op)=\MT.\VerifyMem(\com_A,\ell,\pi)=1$, and halt as in $\Fault$ if
          \begin{itemize}
            \item $\bar\pi$ opens an amount $y^\star<y$ at key $(A,B)$ under $\vroot_\eid$, or
            \item $\MT.\VerifyNM(\vroot_\eid,(A,B),\bar\pi)=1$
          \end{itemize}
  \end{itemize}

  \medskip\noindent
  $\bm{\Fault}$, by the ledger on a successful challenge or on input $\Fault$:
  \begin{itemize}[nosep,leftmargin=1.5em]
    \item \emph{halt}: from then on the ledger answers every input with $\bot$ and rejects closes and challenges, and honest parties return $\bot$ on $\Send$, $\SoftWithdraw$, and $\Close$
    \item output $(\mathsf{Halted},\nxt-1)$ to every party and, with $d_A\coloneq\sum_{(A,x)\in\Req^\Dep}x$ the deposits no close has processed, return $\Bal[A]+d_A$ to every $A$ with $\Bal[A]+d_A>0$ with output $(\Refund,\Bal[A]+d_A)$
  \end{itemize}
\end{breakableprotocol}

\begin{breakableprotocol}{Ideal functionality $\Fvalidity$}{The validity oracle. It reads $\nxt$, $\Bal$, $\Req^\Dep$, $\Req^\FWdr$, $\minreq_\Dep$, $\minreq_\FWdr$, and the registered keys from the ledger, and starts with $S_0=0$.}{fig:validity}
  \noindent$\bm{\Close}$. On $(\Close,\eid,S,\{\bar\st_A\}_A,W,\cnt_\Dep,\cnt_\FWdr,\sigma_\op)$ from $P_\op$, with $S:\cN\times\cN\to\bbN$, $W$ a set of signed soft withdrawal requests, and $\eid=\nxt$:
  \begin{itemize}[nosep,leftmargin=1.5em]
    \item let $\vroot\gets\MT.\Root(S)$
    \item require $\minreq_\Dep\le \cnt_\Dep\le|\Req^\Dep|$ and $\minreq_\FWdr\le \cnt_\FWdr\le|\Req^\FWdr|$, and let $\Dep$ and $\FWdr$ be the first $\cnt_\Dep$ requests of $\Req^\Dep$ and the first $\cnt_\FWdr$ requests of $\Req^\FWdr$
    \item require $\Sig.\Verify(\pk_\op,(\Close,\eid,\vroot,W,\cnt_\Dep,\cnt_\FWdr),\sigma_\op)=1$
    \item for every $A$ with $S[A]\ne0$, parse $\bar\st_A=(\st_A,\sigma_A,\sigma'_A)$, require $\Sig.\Verify(\pk_A,\st_A,\sigma_A)=1$, and require that $\st_A=(\eid,A,\Com(S[A]),\Com(S_{\eid-1}[A]))$
    \item require $q=(\SoftWithdraw,\eid,A,w,\tau)$ with $w>0$ and $\Sig.\Verify(\pk_A,q,\sigma)=1$ for every $(q,\sigma)\in W$, with the requests $q$ pairwise distinct, and let $W[A]$ be the sum of $A$'s amounts $w$ in $W$
    \item let $b[A]\gets\Bal[A]+\sum_{(A,x)\in \Dep}x-\sum_B S[A][B]+\sum_B S[B][A]$ and require $b[A]\ge0$ for every $A$
    \item for every $A$, let $w_A$ be the sum of $A$'s withdrawals in $\FWdr$, let $p[A]\gets\min(b[A],w_A+W[A])$, and let $b'[A]\gets b[A]-p[A]$
    \item store $S_\eid\gets S$ and output $(\Valid,\eid,\vroot,\cnt_\Dep,\cnt_\FWdr,b',p)$ to the ledger
  \end{itemize}
  \medskip\noindent$\bm{\MTOpen}$. On $(\MTOpen,\eid,(A,B))$ from any party with $\eid<\nxt$: if $S_\eid[A][B]=0$, return $\bar\pi\gets\MT.\ProveNM(S_\eid,(A,B))$. Otherwise, return the opening $\bar\pi$ of the leaf $((A,B),S_\eid[A][B])$: the leaf together with $\MT.\ProveMem(S_\eid,i)$, $i$ its position in $S_\eid$.
\end{breakableprotocol}

\begin{remark}
  $\Fvalidity$ can be realized with a committee of $n\geq 2f+1$ validators with at most $f$ malicious. Each validator checks the close (ensuring the requirements above are met) and signs $(\eid,\vroot,\cnt_\Dep,\cnt_\FWdr,b',p)$, and the ledger accepts an $f+1$ certificate, which has an honest signer. A validator checks a close against the matrix $S_{\eid-1}$ that matches the root $\vroot_{\eid-1}$ on the ledger, and does not sign if it does not hold that matrix.
  The operator posts the certificate, and the ledger accepts it only if $\eid=\nxt$ and the ledger has not halted. Until then, nothing the validators checked can change: the queues only grow and lose requests only when a close is posted. A corrupt operator holding a certificate is thus the same as one sending $\Close$ later.
  A corrupt operator may withhold the matrix from users, so the validators, which receive it with the close, answer $\MTOpen$.
\end{remark}

\subsection{Extensions}\label{sec:extensions}
The definition uses the most conservative choices in various parts of the protocol to simplify the modeling. But there are adaptations that an honest operator can use for improved usability. They change only which requests the operator approves. $\Fvalidity$ checks every close as before, so the guarantees against a corrupt operator are unaffected, although \cref{thm:security} is stated for the conservative rule.
The live balance is the most conservative admission rule. An operator may also include any funds the close of $\cur$ is certain to credit: $A$'s deposits among the first $\minreq_\Dep$ requests of $\Req^\Dep$, which the close must process, any later deposits of $A$ that the operator chooses to process in its close, and the payments $A$ has received in $\cur$. Note that the operator need not synchronize on every transfer request if $A$ is not using its full balance. Since both deposits and incoming funds only grow during the epoch, synchronizing periodically may suffice.

\paragraph{Succinct proofs.} In the committee realization of $\Fvalidity$, the operator hands the validators the whole matrix $S_\eid$ and a signed state for every sender. The matrix has an entry for every pair of a sender and a receiver, so it can grow quadratically in the number of accounts, and the validators must store it to answer $\MTOpen$.

If we are willing to relax the security definition, succinct proof systems can reduce the data posted to linear in the number of active accounts. We sketch the necessary modifications below using general-purpose zk-SNARKs, and then describe the relaxation. An interesting direction for future work is a concretely efficient instantiation using algebraic commitments and bespoke signature schemes that allows an operator to prove requisite properties about the close without having to prove the correct execution of heavy cryptographic operations like hash functions and signature verification.

Instead of the validators verifying the checks of \cref{fig:validity}, the operator attaches a succinct proof showing that the close passes verification, and the ledger verifies the proof. Importantly, we do not need zero-knowledge as we do not provide any privacy guarantees. The matrix, the senders' signed states, the soft withdrawals, and the previous matrix $S_{\eid-1}$ form the witness, and the proof checks $S_{\eid-1}$ against the root $\vroot_{\eid-1}$ on the ledger. The close itself carries only the root of the matrix and data that is linear in the number of accounts, namely the balances, the payouts, and the two vectors described next.

What remains is the challenge. In \cref{fig:matrix}, a receiver challenges a close by opening the posted matrix at $(A,B)$, and producing it requires the entire matrix $S_\eid$.
We can instead change the challenge procedure. The variant changes the protocol of \cref{fig:matrix} in four places.
\begin{itemize}
  \item \textbf{Close.} The close additionally posts the total credited to each receiver, $c_\eid[B]\coloneq\sum_A S[A][B]$, and, for every sender $A$ with $S[A]\ne0$, the operator's countersignature on the state it settles. The proof covers both.
  \item \textbf{Challenge.} A receiver $B$ challenges by showing that it holds receipts for $\eid$ from distinct senders whose amounts add up to more than $c_\eid[B]$. The witness is long, as it can hold every receipt that $B$ received in the epoch. $B$ can post the receipts themselves or a succinct proof that it holds them.
  \item \textbf{Sender.} A sender acts only if it signed a state for $\eid$ for which the operator did not return a countersignature. Once the close for $\eid$ is posted, it checks the posted countersignature against that state. If it verifies, the close settled the state, and the sender retrieves receipts for it and provides them to the receiver.
  \item \textbf{Receiver.} A receiver also accepts a receipt for $\eid$ after the close for $\eid$ is posted, as long as its receipts for $\eid$ then add up to at most $c_\eid[B]$. If they add up to more, it challenges.
\end{itemize}
An honest operator will never be challenged successfully, as a receipt from $A$ is for at most the final entry $S[A][B]$.

Observe that a challenge now only compares the \emph{total} credited to a receiver against its receipts, whereas the challenge of \cref{fig:matrix} checks every sender individually. This is where the security definition has to be relaxed. In $\Fclearing$ a close halts whenever $S[A][B]<\Rcpt_\eid[A][B]$ for some sender $A$, which assures an honest receiver that each sender paid what its receipt promises. The variant described above only meets the weaker requirement that a close halts whenever $\sum_A S[A][B]<\sum_A\Rcpt_\eid[A][B]$ for some honest $B$. Under this relaxation, a corrupt operator can omit a payment from $A$ to $B$ without being caught, provided that other senders make up the difference. Depending on how the preconfirmations are used, this may be acceptable.

Importantly, an honest sender is still debited at most what it authorized. As in \cref{fig:matrix}, a corrupt operator may settle a state of an honest sender $A'$ for which it never returned a countersignature, and in the variant it could use such a payment to make up the difference. If $A'$ checks the close, it finds the countersignature and forwards the receipt to $B$. At this point $B$'s receipts add up to more than $c_\eid[B]$, so $B$ challenges and the protocol halts. Thus, provided that honest senders check the close whenever the operator did not return a countersignature for one of their states, the difference can only come from corrupt senders, in which case $B$ is still paid the total it was promised and no honest party loses funds. Finally, since balances and payouts are still posted in full, withdrawals and refunds after a halt work exactly as before.
We do not formalize this variant and leave a full treatment to future work.

\paragraph{Comparison with commit-chains.} Bajillion shares similarities with commit-chains~\cite{EPRINT:KZFMG18}, and in particular with NOCUST-ZKP. In both, a single untrusted operator countersigns the requests of senders and periodically commits the resulting state to the ledger, a commitment only takes effect once it is shown to be valid, and a user holding a countersigned payment that a commitment omits can prove this and halt the system. The two differ in what a sender signs. In NOCUST, a sender signs the list of its individual transfers, and the operator additionally records every transfer in the account of its receiver, together with the total amount that the receiver had received before it. In Bajillion, a sender instead signs the cumulative amount it has paid to each receiver. This has the following consequences.
\begin{itemize}
  \item \textbf{Independent senders.} In NOCUST, the operator must determine the receiver's running total before it countersigns a transfer, and the sender must include this total in its next request. Countersigning thus depends on all earlier transfers to the same receiver, including those of other senders. In Bajillion, the operator checks a request against the state of its sender alone, so requests from different senders are independent even if they pay the same receiver.
        Crucially, this allows the operator to be parallelized up to the number of senders.
  \item \textbf{Smaller closes.} A NOCUST checkpoint accounts for every \emph{transfer} twice, once for its sender and once for its receiver, whereas the matrix $S$ has a single entry for every pair of a sender and a receiver, no matter how many payments were made. When senders make many payments to the same receiver, this reduces the data that $\Fvalidity$ must check and store, and likewise the size of the witness in the variant with succinct proofs sketched above.
\end{itemize}

%% file: security.tex
\section{Security}\label{sec:security}

\begin{theorem}[Security of $\mathsf{Bajillion}$]\label{thm:security}
  Suppose that $\Sig$ is an EUF-CMA-secure signature scheme and $H$ is a collision-resistant hash function. Then $\mathsf{Bajillion}$ (\cref{fig:matrix}) securely emulates $\Fclearing$ (\cref{fig:func}) in the $\Fvalidity$-hybrid model against static malicious corruption of any set of users and, optionally, the operator, with an honest ledger, in the sense of \cref{def:emulation}.
\end{theorem}

\begin{proof}
  Let $\adv$ be a PPT adversary that statically corrupts a set $M$ of users and possibly $P_\op$. As corruption is static, the operator is honest throughout the execution or corrupt throughout it, and we treat the two cases separately.

  \paragraph{Simulator.} $\Sim$ runs the real protocol internally with $\adv$. It plays the ledger, $\Fvalidity$, and every honest party, with keys it samples itself, and follows the protocol faithfully. $\Fclearing$ leaks every honest input to $\Sim$, as we demand no privacy for user inputs.
  \begin{itemize}[nosep,leftmargin=1.5em]
    \item \emph{Honest inputs.} Forward the input of an honest user to the corresponding simulated user, a $\Close$ input of an honest operator to the simulated operator, and an input of $P_L$ to the simulated ledger.
    \item \emph{Corrupt user, honest operator.} When the simulated operator accepts a signed state of a corrupt $A$ with batch $X$, $\Sim$ sends $(\Send,A,X,\hlt{\cur},\mathsf{false})$ to $\Fclearing$ on $A$'s behalf, once for each accepted state. For each soft withdrawal $(\SoftWithdraw,\eid,A,w,\tau)$ of a corrupt $A$ that the operator accepts, $\Sim$ sends $(\SoftWithdraw,A,w,\eid)$ on $A$'s behalf.
    \item \emph{Close, corrupt operator.} When the simulated ledger posts a close of the corrupt operator, $\Sim$ sends $(\Close,\cnt_\Dep,\cnt_\FWdr,S,b')$ to $\Fclearing$ on $P_\op$'s behalf. Here $S$, $\cnt_\Dep$, and $\cnt_\FWdr$ are the close's matrix and counts, and $b'$ are the new balances that $\Fvalidity$ computed.
    \item \emph{Receipt.} When a simulated honest $B$ accepts a receipt from $A$ for epoch $\eid$ with amount $y$, and $A$ or $P_\op$ is corrupt, send $(\Release,A,B,\eid,y)$.
    \item \emph{Halt.} When the simulated ledger halts on a challenge posted by a corrupt party, and $\Fclearing$ has not already halted, send $\Fault$. For a challenge posted by an honest receiver, $\Fclearing$ raises the halt itself, in $\Close$.
  \end{itemize}
  So $\Sim$ sends $\Send$ and $\SoftWithdraw$ only under an honest operator, and $\Close$ only under a corrupt one. We show that it sends $\Fault$ only under a corrupt operator.

  \paragraph{Hybrids.}
  \begin{hybrids}[nosep,leftmargin=2.5em]
    \item The real execution of $\mathsf{Bajillion}$ with $\adv$ and $\cZ$.
    \item\label{hyb:forge} As before, but abort if an honest party, the ledger, or $\Fvalidity$ accepts a signature under an honest party's key on a message that party never signed. This is negligible by the EUF-CMA security of $\Sig$.
    \item\label{hyb:collide} As before, but abort if two distinct sorted lists computed in the execution have the same root, or if an honest party, the ledger, or $\Fvalidity$ accepts, under the root of a sorted list, a membership proof opening a value the list does not hold at that key or a non-membership proof for a key the list contains. This is negligible by the collision resistance of $H$.
    \item\label{hyb:run} As before, but also run $\Fclearing$ on the honest inputs and on $\Sim$'s commands, and ignore its outputs.
    \item\label{hyb:ideal} As before, but honest parties output what $\Fclearing$ outputs to them. Up to the aborts, this is the ideal execution with $\Sim$.
  \end{hybrids}
  Hybrids \ref{hyb:run} and \ref{hyb:ideal} are identical if the invariants below hold in \ref{hyb:run}, by \ref{inv:out}.

  \paragraph{Invariants.} A shaded box marks a variable of the simulated protocol. $\hlt{\nxt}$, $\hlt{\Bal}$, $\hlt{\Req^\Dep}$, $\hlt{\Req^\FWdr}$, $\hlt{\minreq_\Dep}$, and $\hlt{\minreq_\FWdr}$ are the simulated ledger's. $\hlt{\cur}$, $\hlt{\Auth_\eid}$, $\hlt{W_\eid}$, and $\hlt{\live}$ are the simulated operator's. $\hlt{S_\eid}$ is the simulated $\Fvalidity$'s, and $\hlt{\Rcpt^B_\eid}$ is the simulated $B$'s. Unmarked variables are those of $\Fclearing$, except for three that exist only in the protocol: the payments maps $\cP_{A,e}$ and the pending request of a simulated honest $A$, and the roots $\vroot_\eid$.

  The following hold after every event.
  \begin{enumerate}[nosep,leftmargin=2em,label=(I\arabic*)]
    \item\label{inv:epoch} $\cur=\hlt{\nxt}$, and if $P_\op$ is honest, $\hlt{\cur}=\hlt{\nxt}$. $\Fclearing$ has halted iff the simulated ledger has.
    \item\label{inv:dw} $\Bal=\hlt{\Bal}$, $\Req^\Dep=\hlt{\Req^\Dep}$, $\Req^\FWdr=\hlt{\Req^\FWdr}$, $\minreq_\Dep=\hlt{\minreq_\Dep}$, and $\minreq_\FWdr=\hlt{\minreq_\FWdr}$. For every $e<\cur$, $\Fclearing$ closed $e$ with the matrix $\hlt{S_e}$, where $\hlt{S_0}=0$.
    \item\label{inv:auth} Unless halted, the following hold:
          \begin{itemize}[leftmargin=*]
            \item If $P_\op$ is honest: $\Auth_e=\hlt{\Auth_e}$ for every $e$, $\hlt{S_e}=\hlt{\Auth_e}$ for every $e<\cur$, and $\Pipe_e=\Carry_e=0$ for every $e$. $\Wdr_\cur[A]$ is the sum of $A$'s amounts in $\hlt{W_\cur}$ for every $A$. $\live(A)=\hlt{\live}(A)$ for every $A$. $\cP_{A,e}=\hlt{\Auth_e}[A]$ for every honest $A$ and every $e$.
            \item If $P_\op$ is corrupt: for every honest $A$ and every $e\ge\cur$, the amounts of the distinct requests $(\SoftWithdraw,e,A,\cdot,\cdot)$ that $A$ has signed sum to at most $\Wdr_e[A]$.
          \end{itemize}
    \item\label{inv:rcpt} Unless halted, $\Rcpt_e[A][B]=\hlt{\Rcpt^B_e}[A]$ for every honest $B$, every $A$, and every $e\ge\cur$.
    \item\label{inv:out} For every honest party, the sequence of outputs $\Fclearing$ has delivered to it is identical to the sequence of outputs its simulated copy has produced.
  \end{enumerate}
  For a batch $X=\{(B_i,x_i)\}_i$, let $X[B]$ be the sum of the $x_i$ with $B_i=B$. Adding $X$ to a payments map adds $X[B]$ to the entry of every $B$.

  \paragraph{Induction.} We show by induction over events that \ref{inv:epoch}--\ref{inv:rcpt} hold, that $\Fclearing$ accepts every command of $\Sim$, and that $\Fclearing$ and the simulated honest parties give every honest party the same outputs in the same order, which preserves \ref{inv:out}.
  \begin{itemize}[nosep,leftmargin=1.5em]
    \item \emph{Base case.} Initially $\Fclearing$ and the simulated parties hold the same balances $\Init$ and the epoch $1$, the queues are empty, all tables are zero, and nothing has been signed or output.
    \item \emph{Inductive step.} Assume that this holds before an event. The rest of the proof shows that it holds after the event, first for the events that do not depend on the operator, and then for an honest operator (Case 1) and a corrupt operator (Case 2). Before a halt, honest parties output only $\mathsf{Received}$, $\mathsf{Paid}$, $\mathsf{Finalized}$, and $\bot$. Outputs to corrupt parties go to $\Sim$, which ignores them.
  \end{itemize}

  \paragraph{Events in both cases.} The following steps do not depend on the operator.
  \begin{itemize}[nosep,leftmargin=1.5em]
    \item \emph{Deposits and forced withdrawals.} $\Fclearing$ and the simulated ledger append the input's pair to the same queue, which preserves \ref{inv:dw}.
    \item \emph{Ledger fault.} $\Sim$ forwards the input $\Fault$ of $P_L$ to the simulated ledger, and both halt.
    \item \emph{Halt.} If $\Fclearing$ and the simulated ledger halt in the same event, both output $(\mathsf{Halted},\cur-1)$ and then $(\Refund,\Bal[A]+d_A)$ for the same $A$ and $d_A$, by \ref{inv:epoch} and \ref{inv:dw}. Afterwards $\Fclearing$ returns $\bot$ on every input. The simulated ledger rejects deposits, withdrawals, closes, and challenges. Simulated honest parties return $\bot$ on $\Send$, $\SoftWithdraw$, and $\Close$, and produce no other output, as receivers reject receipts once the ledger has halted. So every later event gives the same outputs and changes no invariant.
    \item \emph{Receipts.} Let an honest $B$ accept a receipt from $A$ for epoch $\eid$ with amount $y$, where $A$ or $P_\op$ is corrupt. Then $y>\hlt{\Rcpt^B_\eid}[A]$, and $\eid\ge\hlt{\nxt}=\cur$ by \ref{inv:epoch}. Each case shows that $\Fclearing$ accepts $\Sim$'s $(\Release,A,B,\eid,y)$. As $y>\Rcpt_\eid[A][B]$ by \ref{inv:rcpt}, $\Fclearing$ and the simulated $B$ both set their entry to $y$ and output $(\mathsf{Received},\eid,A,y)$ to $B$, and nothing else.
    \item \emph{Other events.} Any other event leaves the variables in the invariants unchanged, and produces either no output or the same $\bot$ in both executions. Examples are an input that violates the requirements of its handler, a message between corrupt parties, a $\bot$ that does not yet drop a pending request, an epoch announcement, an $\MTOpen$ query, and a rejected receipt, close, or challenge.
  \end{itemize}

  \funcsep

  \paragraph{Case 1: honest operator.} Let $P_\op$ be honest. By \ref{inv:auth}, the simulated operator and $\Fclearing$ hold the same maps, and the two move in lockstep. The two claims below assume that \ref{inv:epoch}--\ref{inv:rcpt} have held so far and that no halt has occurred.

  \begin{claim}[honest operator]\label{clm:honest-op}
    If $P_\op$ is honest, then:
    \begin{enumerate}[nosep,leftmargin=2em,label=(\alph*)]
      \item\label{cl:close} If the operator gets a $\Close$ input and its own checks pass, that is, $\hlt{\nxt}=\hlt{\cur}$ and both counts lie within the ledger's bounds, then $\Fvalidity$ accepts the operator's close and posts it to the ledger.
      \item\label{cl:rcpt} In $\Fclearing$, $\Rcpt_e[A][B]\le\Auth_e[A][B]$ for all $A$, $B$, and every $e\ge\cur$.
      \item\label{cl:op-rcpt} If a receipt $(\bar\st,(B,y),\pi)$ with $\st=(e,A,\com,\com')$ carries verifying signatures of $A$ and $P_\op$ on $\st$ and a membership proof for $(B,y)$ that verifies under $\com$, then $0<y\le\hlt{\Auth_e}[A][B]$.
    \end{enumerate}
  \end{claim}

  \begin{proof}
    We first note two facts about the honest operator.
    \begin{itemize}[leftmargin=1.5em]
      \item \emph{First fact.} Every state $(e,A,\Com(\cP),\com')$ it has countersigned has $e\le\hlt{\cur}$, $\com'=\Com(\hlt{\Auth_{e-1}}[A])$, and $\cP\le\hlt{\Auth_e}[A]$, and the last one it countersigned for $A$ in $e$ commits to $\hlt{\Auth_e}[A]$. This holds because the operator countersigns only states for $\hlt{\cur}$, after checking $\com'$, and sets $\hlt{\Auth_e}[A]$ to the state's map when it countersigns. It never changes $\hlt{\Auth_{e-1}}$ once $\hlt{\cur}=e$, and never decreases $\hlt{\Auth_e}[A]$.

      \item \emph{Second fact.} $\hlt{\live}(A)=\hlt{\Bal}[A]-\sum_B\hlt{\Auth_\cur}[A][B]\ge0$ for every $A$. We show this by induction over the steps of the operator.
            \begin{itemize}[leftmargin=1.5em]
              \item \emph{Base case.} Initially $\hlt{\live}=\hlt{\Bal}=\Init$ and $\hlt{\Auth_\cur}=0$.
              \item \emph{Inductive step.} The operator changes these variables only when it accepts a request and when it closes. When it accepts a request of $A$, it lowers $\hlt{\live}(A)$ by $\sum_ix_i\le\hlt{\live}(A)$, and $\hlt{\Auth_\cur}[A]$ grows by the same amount. At a close it resets $\hlt{\live}$ to $b'\ge0$ and starts the new epoch with $\hlt{\Auth_\cur}=0$, and by \ref{cl:close} the ledger sets $\hlt{\Bal}$ to the same $b'$ in the same event.
            \end{itemize}
    \end{itemize}

    For \ref{cl:close}, let $\eid=\hlt{\cur}$, which is $\hlt{\nxt}$ by \ref{inv:epoch}. The operator checked the counts against the ledger's queues and bounds, which $\Fvalidity$ reads. The close carries $S=\hlt{\Auth_\eid}$, the operator's signature, and the soft withdrawals for $\eid$ that the operator verified, which are pairwise distinct and have positive amounts. For every $A$ with $S[A]\ne0$, it also carries the last state the operator countersigned for $A$ in $\eid$, with $A$'s signature. By the first fact, that state is $(\eid,A,\Com(S[A]),\Com(\hlt{\Auth_{\eid-1}}[A]))$, and $\hlt{\Auth_{\eid-1}}=\hlt{S_{\eid-1}}$ by \ref{inv:auth}. By the second fact, $b[A]=\hlt{\live}(A)+d_A+\sum_BS[B][A]\ge0$. So $\Fvalidity$ accepts the close and posts it, and the ledger applies it, as it has not halted.

    For \ref{cl:rcpt}, $\Rcpt_e[A][B]$ is set only by $\Send$, to $\Auth_e[A][B]$, or by $\Release$, to some $y\le\Auth_e[A][B]$, as $P_\op$ is honest and $\Pipe=0$. $\Auth_e$ never decreases.

    For \ref{cl:op-rcpt}, $P_\op$ countersigned $\st$ by \ref{hyb:forge}. So by the first fact, $\com=\Com(\cP)$ for some $\cP$ with $\cP[B]\le\hlt{\Auth_e}[A][B]$. By \ref{hyb:collide}, $y=\cP[B]$, which is positive as leaves hold only nonzero amounts.
  \end{proof}

  \begin{claim}[mirroring]\label{clm:mirror}
    In each step below, $\Fclearing$ accepts the commands of $\Sim$, and the simulated parties change their variables as $\Fclearing$ changes its own and produce the same outputs. This preserves \ref{inv:epoch}--\ref{inv:auth}.
    \begin{enumerate}[nosep,leftmargin=2em,label=(\alph*)]
      \item\label{cl:m-send} The operator accepts a signed state of $A$ with batch $X$, and $\Fclearing$ adds $X$ to $\Auth_\cur[A]$. For an honest $A$, $\Fclearing$ does so on $A$'s own $\Send$ input. For a corrupt $A$, it does so on $\Sim$'s $(\Send,A,X,\cur,\mathsf{false})$.
      \item\label{cl:m-soft} A soft withdrawal $(\SoftWithdraw,\eid,A,w,\tau)$ reaches the operator, which adds it to $\hlt{W_\cur}$ iff $\Fclearing$ adds $w$ to $\Wdr_\cur[A]$.
      \item\label{cl:m-close} $\Fclearing$ gets $(\Close,\cnt_\Dep,\cnt_\FWdr)$. The operator's checks pass iff $\Fclearing$'s do, and otherwise both return $\bot$. If they pass, $\Fclearing$ closes $\cur$, and $\Fvalidity$ posts the operator's close to the ledger, which applies it.
    \end{enumerate}
  \end{claim}

  \begin{proof}
    By \ref{inv:epoch} and \ref{inv:auth}, $\hlt{\cur}=\hlt{\nxt}=\cur$ and $\hlt{\live}=\live$. The operator answers every request when it is made, so no honest sender has a pending request.

    For \ref{cl:m-send}, the operator only accepts states for $\cur$ with $m\ge1$, every $B_i\in\cN$, every $x_i>0$, the $B_i$ pairwise distinct, and $\sum_ix_i\le\live(A)$. So $\Fclearing$ accepts $\Sim$'s command for a corrupt $A$, and sets no receipt. The operator sets $\hlt{\Auth_\cur}[A]$ to $\hlt{\Auth_\cur}[A]+X$, after checking that the state commits to this map, and lowers $\hlt{\live}(A)$ by $\sum_ix_i$. Adding $X$ to $\Auth_\cur[A]$ lowers $\live(A)$ by the same amount.

    For \ref{cl:m-soft}, the operator adds a request iff it is new, signed by $A$, for $\eid=\cur$, and has $w>0$. If $A$ is corrupt, $\Sim$ then sends $(\SoftWithdraw,A,w,\eid)$, and $\Fclearing$ adds $w$ to $\Wdr_\cur[A]$. If $A$ is honest, the request is new, as its tag is fresh, and comes with $A$'s input to $\Fclearing$, which adds $w$ iff $\eid=\cur$. The adversary cannot send the operator a request of an honest $A$, as $A$ signed it by \ref{hyb:forge} and sent it over a private channel.

    For \ref{cl:m-close}, let $\eid=\cur$. The operator's check $\hlt{\nxt}=\hlt{\cur}$ holds, and both check the counts against the same queues and bounds, by \ref{inv:dw}. Let the checks pass. By Claim~\ref{clm:honest-op}\ref{cl:close}, $\Fvalidity$ accepts the operator's close, with $S=\hlt{\Auth_\eid}$ and $W=\hlt{W_\eid}$, and posts it. Its $b$ and $p$ are $\Fclearing$'s, as both use the same balances and requests by \ref{inv:dw}, and $S=\Auth_\eid$ and $W[A]=\Wdr_\eid[A]$ by \ref{inv:auth}. So the ledger and $\Fclearing$ make the same changes to the balances, queues, and counters, give the same $\mathsf{Paid}$ and $\mathsf{Finalized}$ outputs, and advance $\hlt{\nxt}$ and $\cur$ to $\eid+1$. $\Fclearing$ closes $\eid$ with $\Auth_\eid$, which $\Fvalidity$ stores as $\hlt{S_\eid}$, and sets $\Carry_\eid=0$, as $\Pipe_\eid=\Carry_{\eid-1}=0$. The operator already set $\hlt{\cur}\gets\eid+1$ and $\hlt{\live}\gets b'$ when it sent the close. The new $\Auth_\cur$, $\Wdr_\cur$, $\hlt{\Auth_\cur}$, and $\hlt{W_\cur}$ are zero or empty, so $\live=\Bal=b'=\hlt{\live}$. No honest sender changes its payments map, as none has a pending request.
  \end{proof}

  We now go through all possible events.

  \paragraph{Send by an honest party.} Let an honest $A$ get $(\Send,A,X,\eid,\mathsf{pipe})$. $A$ has no pending request, as the honest operator answers every state of a request as soon as it arrives. If $\cP_{A,e}\neq0$ for some $e>\eid$, $A$ ignores the input, and so does $\Fclearing$, as $\Auth_e[A]=\cP_{A,e}$ by \ref{inv:auth}. Otherwise $A$ signs a state for $\eid$, and an additional one for $\eid+1$ if $\mathsf{pipe}=\mathsf{true}$, and sends the request. The operator handles each state on its own and answers $\bot$ for every state for an epoch other than $\hlt{\cur}=\cur$. So it countersigns at most one state of the request. If none of the states is for $\cur$, $A$ receives a $\bot$ for each of them and drops the request, and $\Fclearing$ does nothing.

  Otherwise one state is for $\cur$, namely $(\cur,A,\Com(\cP'),\Com(\cP_{A,\cur-1}))$ with $\cP'=\cP_{A,\cur}+X$. By \ref{inv:auth}, $\cP_{A,\cur-1}=\hlt{\Auth_{\cur-1}}[A]$ and $\cP'=\hlt{\Auth_\cur}[A]+X$, so the operator's signature and commitment checks pass. So it accepts the state iff $\sum_ix_i\le\hlt{\live}(A)$, which is $\Fclearing$'s test by \ref{inv:auth}. If the test fails, $A$ drops the request, and $\Fclearing$ does nothing. Otherwise $A$ accepts the operator's $(\cur,\sigma_\op)$ and sets $\cP_{A,\cur}\gets\cP'$, and $\Fclearing$ adds $X$ to $\Auth_\cur[A]$.
  With Claim~\ref{clm:mirror}\ref{cl:m-send}, this preserves \ref{inv:epoch}--\ref{inv:auth}.

  $A$ then sends each $B_i$ a receipt for $(B_i,\cP'[B_i])$. An honest $B_i$ accepts it: both signatures and the proof verify, and $\cP'[B_i]=\Auth_\cur[A][B_i]$ exceeds $\Rcpt_\cur[A][B_i]=\hlt{\Rcpt^{B_i}_\cur}[A]$, by Claim~\ref{clm:honest-op}\ref{cl:rcpt} as $x_i>0$, and by \ref{inv:rcpt}. $\Fclearing$ and the simulated $B_i$ both set that entry to $\cP'[B_i]$ and output $(\mathsf{Received},\cur,A,\cP'[B_i])$ to $B_i$.

  Since every accepted send of $A$ that pays $B$ also updates $B$'s entry, $\hlt{\Rcpt^B_\eid}[A]=\hlt{\Auth_\eid}[A][B]$ outside this event. So any other receipt from $A$ fails $B$'s check, as its amount is at most $\hlt{\Auth_\eid}[A][B]$ by Claim~\ref{clm:honest-op}\ref{cl:op-rcpt}.

  \paragraph{Corrupt senders and soft withdrawals.} When the operator accepts a state of a corrupt $A$, $\Sim$ sends $\Send$ on $A$'s behalf, and Claim~\ref{clm:mirror}\ref{cl:m-send} applies. Claim~\ref{clm:mirror}\ref{cl:m-soft} covers soft withdrawals of honest and corrupt users.

  \paragraph{Close.} $\Fvalidity$ takes closes only from $P_\op$, so it accepts only the operator's own closes. Claim~\ref{clm:mirror}\ref{cl:m-close} covers them, and \ref{inv:rcpt} still holds, as $\cur$ only grows. After the close of $\eid$, $\Fclearing$ does not halt, as $S=\Auth_\eid\ge\Rcpt_\eid$ by Claim~\ref{clm:honest-op}\ref{cl:rcpt}. No honest $B$ challenges the close either. $B$ would challenge only if the posted matrix had $S[A][B]<\hlt{\Rcpt^B_\eid}[A]$ for some $A$, and this entry is $\Rcpt_\eid[A][B]$, as \ref{inv:rcpt} held before the close.

  \paragraph{Receipts from a corrupt sender.} Let an honest $B$ accept a receipt from a corrupt $A$ for epoch $\eid$ with amount $y$. By Claim~\ref{clm:honest-op}\ref{cl:op-rcpt} and \ref{inv:auth}, $y\le\Auth_\eid[A][B]$, and $\Pipe=0$. So $\Fclearing$ accepts $\Sim$'s $\Release$, and the receipt step of both cases applies.

  \paragraph{Challenges by corrupt parties.} No challenge succeeds. Let a challenge against the posted close for $\eid$ carry a verifying receipt for $(B,y)$ from $A$. The close has matrix $S=\hlt{S_\eid}=\hlt{\Auth_\eid}$ by \ref{inv:auth}. So $0<y\le S[A][B]$ by Claim~\ref{clm:honest-op}\ref{cl:op-rcpt}. By \ref{hyb:collide}, no opening below $y$ at $(A,B)$ under $\vroot_\eid=\MT.\Root(S)$ verifies, and neither does a non-membership proof for $(A,B)$. So $\Sim$ never sends $\Fault$. With the close step, this shows that an honest operator is never halted by a challenge. $\Fclearing$ and the simulated ledger both halt only on the input $\Fault$ of $P_L$, which preserves \ref{inv:epoch}.

  \funcsep

  \paragraph{Case 2: corrupt operator.} Let $P_\op$ be corrupt. The operator is then part of the adversary, and $\Sim$ does not simulate it, so no invariant refers to its variables. Instead, \ref{inv:auth} and the claim below relate what honest users sign to the maps of $\Fclearing$. The claim assumes that \ref{inv:epoch}--\ref{inv:rcpt} have held so far and that no halt has occurred.

  \begin{claim}[honest sender, corrupt operator]\label{clm:sender}
    Let $P_\op$ be corrupt, $A$ honest, and $B$ any account. The requests of $A$ are the $\Send$ inputs it does not ignore. We say that $A$ \emph{books} a request in epoch $e$ when it receives a valid answer $(e,\sigma_\op)$ for it, or finds it in the close of $e$ with its $\MTOpen$ check. Then:
    \begin{enumerate}[nosep,leftmargin=2em,label=(\alph*)]
      \item\label{cl:count} $A$ books each request at most once, in its epoch $\eid$ or, if the request is pipelined, in $\eid+1$. At any time, $\cP_{A,e}$ is the sum of the batches that $A$ has booked in $e$.
      \item\label{cl:record} $\Fclearing$ records each request of $A$ when $A$ makes it, unless every epoch the request may settle in is already closed. It adds the batch of a plain request for $\eid$ to $\Auth_\eid[A]$. It adds the batch of a pipelined request for $\eid$ to $\Pipe_\eid[A]$ if $\eid$ is still open, and to $\Auth_{\eid+1}[A]$ otherwise.
      \item\label{cl:debit} If $\Fvalidity$ accepts a close for $\cur$ with matrix $S$, then $S[A][B]\le\Auth_\cur[A][B]+\Carry_{\cur-1}[A][B]+\Pipe_\cur[A][B]$.
      \item\label{cl:receipt} If $B$ is honest and accepts a receipt from $A$ for epoch $\eid$ with amount $y$, then $\eid\ge\cur$ and $y\le\Auth_\eid[A][B]+\Pipe_{\eid-1}[A][B]+\Pipe_\eid[A][B]$.
    \end{enumerate}
  \end{claim}

  \begin{proof}
    For \ref{cl:count}, $A$ books only its pending request, in $\eid$ or, if the request is pipelined, in $\eid+1$, after which it is no longer pending. Booking it in $e$ sets $\cP_{A,e}$ to the map in its state for $e$, which is $\cP_{A,e}+X$, as $\cP_{A,e}$ has not changed since $A$ made the request.

    For \ref{cl:record}, an epoch is closed once its close is posted, and open before. So $e$ is open iff $e\ge\cur$, as $\cur=\hlt{\nxt}$ by \ref{inv:epoch}. Let some epoch that the request may settle in be open when $A$ makes it. If the request is plain, or pipelined with $\eid$ open, then $\eid\ge\cur$. If it is pipelined and $\eid$ is closed, then $\eid<\cur\le\eid+1$, so $\eid=\cur-1$. These are the three cases of $\Fclearing$'s $\Send$ for an honest $A$ and a corrupt $P_\op$.

    For \ref{cl:receipt} and \ref{cl:debit}, the idea is as follows. A receipt and a close both carry a state that $A$ signed, and such a state includes each batch of $A$ once. The previous-epoch commitment stops a pipelined batch that was settled in $\eid-1$ from being settled again in $\eid$.

    \medskip\noindent\textbf{Notation.} Let $X_r$ be the batch of a request $r$, $\eid_r$ its epoch, and $t_r$ the time at which $A$ makes it. Let $R_\eid(t)$ be the set of requests that $A$ books in $\eid$ before time $t$, and $\cP_{A,\eid}(t)$ the value of $\cP_{A,\eid}$ at $t$. By \ref{cl:count}, $\cP_{A,\eid}(t)=\sum_{r\in R_\eid(t)}X_r$. Let $Q_\eid$ be the set of pipelined requests for $\eid$ that $A$ makes while $\eid$ is open. By \ref{cl:record}, their batches are in $\Pipe_\eid[A]$.

    \begin{itemize}[leftmargin=1.5em]

      \item \textbf{Recorded requests.} A request may settle in $\eid$ if it is for $\eid$, or pipelined for $\eid-1$. Let $A$ make such a request while $\eid$ is open. By \ref{cl:record}, a plain request for $\eid$ is recorded in $\Auth_\eid[A]$, and a pipelined one in $\Pipe_\eid[A]$. A pipelined request for $\eid-1$ is in $Q_{\eid-1}$ if $A$ made it while $\eid-1$ was open, and is recorded in $\Auth_\eid[A]$ otherwise. Distinct requests are distinct $\Send$ inputs, and $\Auth$ and $\Pipe$ never decrease. So take any set of requests that may each settle in $\eid$ and that $A$ made while $\eid$ was open. Summing over this set,
            \begin{equation}\label{eq:record}
              \textstyle\sum_{r}X_r[B]\le\Auth_\eid[A][B]+\Pipe_\eid[A][B]+\sum_{r\in Q_{\eid-1}}X_r[B].
            \end{equation}
            Inputs that $A$ ignores, which $\Fclearing$ may also record as it does not see $A$'s pending request or booked maps, only add to the right-hand side.

      \item \textbf{Signed states.} Consider a state $(\eid,A,\com,\com')$ with a verifying signature of $A$, where $\eid\ge\cur$, as carried by a receipt or a close. By \ref{hyb:forge}, $A$ signed it, and $A$ signs a state only when it makes a request. Let $r$ be a request for which $A$ signed this state. It need not be $A$'s latest request, or one that the operator answered, as a corrupt operator may use any state that $A$ has signed. By the sender's code, $\com$ commits to $\cP_{A,\eid}(t_r)+X_r$, which is the sum of the batches of $r$ and of the requests in $R_\eid(t_r)$, and $\com'$ commits to $\cP_{A,\eid-1}(t_r)$. Each of these requests may settle in $\eid$, and $A$ made it by $t_r$, so while $\eid$ was open, as $\eid$ is open now. So \eqref{eq:record} applies to them.

      \item \textbf{Receipt bound \ref{cl:receipt}.} $B$ checks that $\eid\ge\hlt{\nxt}$, which is $\cur$ by \ref{inv:epoch}, that $A$'s signature on the state of the receipt verifies, and that a membership proof for $(B,y)$ verifies under $\com$. By \ref{hyb:collide}, $y$ is the entry of $B$ in $\cP_{A,\eid}(t_r)+X_r$. The batches of the requests in $Q_{\eid-1}$ are in $\Pipe_{\eid-1}[A]$, so \eqref{eq:record} gives the bound.

      \item \textbf{Debit bound \ref{cl:debit}.} The bound is trivial if $S[A]=0$, so let $S[A]\ne0$. Then $\Fvalidity$ verified $A$'s signature on a state for $\cur$ that commits to $S[A]$ and to $\hlt{S_{\cur-1}}[A]$. As above, $A$ signed this state for some request $r$. So by \ref{hyb:collide}, $S[A]=\cP_{A,\cur}(t_r)+X_r$ and $\hlt{S_{\cur-1}}[A]=\cP_{A,\cur-1}(t_r)$. That is, $S[A]$ is the sum of the batches of $r$ and of the requests in $R_\cur(t_r)$, which $A$ booked in $\cur$ before it made $r$.

            At most one of these requests carries over from $\cur-1$, that is, is in $Q_{\cur-1}$. To see this, order these requests by the time at which $A$ makes them. The request $r$ is the latest, as $A$ booked the others before it made $r$. Let $r^\star$ be the earliest. If $r^\star=r$, there are no others. Otherwise $A$ books $r^\star$ in $\cur$. As $A$ makes no request while another one is pending, it makes all the others after it has booked $r^\star$. From then on $\cP_{A,\cur}\neq0$, so $A$ ignores every $\Send$ input for an epoch before $\cur$. So none of the others are for $\cur-1$, and only $r^\star$ can carry over.

            Those that do not carry over are recorded in $\Auth_\cur[A]$ or $\Pipe_\cur[A]$, so by \eqref{eq:record} they pay $B$ at most $\Auth_\cur[A][B]+\Pipe_\cur[A][B]$ in total. If $r^\star$ carries over, it remains to show that $X_{r^\star}[B]\le\Carry_{\cur-1}[A][B]$.
            For this we show that $A$'s map for $\cur-1$ is the same at $t_{r^\star}$ and at $t_r$. This is clear if $r^\star=r$. Otherwise, as shown above, $A$ books $r^\star$ in $\cur$, and every request that it makes afterwards is for $\cur$ or a later epoch. So $A$ books nothing in $\cur-1$ after it makes $r^\star$. Hence $\hlt{S_{\cur-1}}[A]=\cP_{A,\cur-1}(t_r)=\cP_{A,\cur-1}(t_{r^\star})$, which is the condition of the carry bound below for the epoch $\cur-1$. The carry bound then gives $X_{r^\star}[B]\le\Carry_{\cur-1}[A][B]$.

      \item \textbf{Carry bound.} We claim the following. Let $\eid<\cur$, and let $r^\star$ be a request in $Q_\eid$ such that the close of $\eid$ settled for $A$ what $A$ had booked in $\eid$ when it made $r^\star$, that is, $\hlt{S_\eid}[A]=\cP_{A,\eid}(t_{r^\star})$. Then
            \begin{equation}\label{eq:carry}
              X_{r^\star}[B]\le\Carry_\eid[A][B].
            \end{equation}
            We prove the claim by induction on $\eid$.
            \begin{itemize}[leftmargin=1.5em]
              \item \emph{Base case.} For $\eid=0$, there is no such request, as $Q_0$ is empty.

              \item \emph{Inductive step.} Let $\eid\ge1$, and assume the claim for $\eid-1$. By \ref{inv:dw}, $\Fclearing$ closed $\eid$ with $\hlt{S_\eid}$, so
                    \[\Carry_\eid[A][B]=\Pipe_\eid[A][B]-\max\bigl(0,\ \hlt{S_\eid}[A][B]-\Auth_\eid[A][B]-\Carry_{\eid-1}[A][B]\bigr).\]
                    So the claim amounts to two inequalities. The first is $X_{r^\star}[B]\le\Pipe_\eid[A][B]$, which holds as $r^\star$ is recorded in $\Pipe_\eid[A]$. The second is
                    \[\hlt{S_\eid}[A][B]+X_{r^\star}[B]\le\Auth_\eid[A][B]+\Carry_{\eid-1}[A][B]+\Pipe_\eid[A][B].\]
                    Without $X_{r^\star}[B]$, this is the debit bound for the close of $\eid$. We show that the proof of the debit bound still goes through with $r^\star$ added.

                    If $\hlt{S_\eid}[A]=0$, the second inequality follows from the first. Otherwise, recall the proof of the debit bound for the close of $\eid$, which $\Fvalidity$ accepted while $\eid$ was the current epoch. The close carried a state that $A$ signed for some request $r$, and $\hlt{S_\eid}[A]$ is the sum of the batches of $r$ and of the requests in $R_\eid(t_r)$. Only the earliest of these requests, call it $r'$, can carry over from $\eid-1$. The proof bounds these requests in two parts.
                    \begin{itemize}[leftmargin=1.5em]
                      \item The requests that do not carry over from $\eid-1$ pay $B$ at most $\Auth_\eid[A][B]+\Pipe_\eid[A][B]$ in total, by \eqref{eq:record}. This remains true with $r^\star$ added. It is not one of these requests: if it were $r$, or if $A$ had booked it in $\eid$ before $t_r$, then $\hlt{S_\eid}[A]$ would include the batch of $r^\star$ on top of $\cP_{A,\eid}(t_{r^\star})$, and could not equal it. And like them, it may settle in $\eid$, $A$ made it while $\eid$ was open, and it does not carry over from $\eid-1$.

                      \item If $r'$ carries over from $\eid-1$, then $\hlt{S_{\eid-1}}[A]=\cP_{A,\eid-1}(t_{r'})$, and the proof uses the carry bound for $\eid-1$. This is our induction hypothesis, so $X_{r'}[B]\le\Carry_{\eid-1}[A][B]$.
                    \end{itemize}
                    Together, the two parts give the second inequality.\qedhere
            \end{itemize}
    \end{itemize}
  \end{proof}

  We now go through the events of this case.

  \paragraph{Sends of honest senders.} The requests of an honest $A$, the answers it accepts, and the requests it drops change only its payments maps and $\Fclearing$'s $\Auth$ and $\Pipe$, as Claim~\ref{clm:sender} describes. No invariant refers to these variables, and these steps produce no output. The receipts that $A$ sends are treated below.

  \paragraph{Soft withdrawals.} Let an honest $A$ get the input $(\SoftWithdraw,A,w,\eid)$. The request that it signs has a fresh tag, so it is distinct from its earlier ones. If $\eid\ge\cur$, $\Fclearing$ adds $w$ to $\Wdr_\eid[A]$, which preserves \ref{inv:auth}. If $\eid<\cur$, $\Fclearing$ does nothing, and \ref{inv:auth} does not cover $\eid$, now or later, as $\cur$ only grows.

  \paragraph{Close.} Let $\Fvalidity$ accept a close for $\eid=\hlt{\nxt}=\cur$ with matrix $S$, soft withdrawals $W$, and counts $\cnt_\Dep$ and $\cnt_\FWdr$. Let $W[A]$ be the sum of $A$'s amounts in $W$.
  \begin{itemize}[leftmargin=1.5em]
    \item \emph{$\Fclearing$ accepts the close.} Every check of $\Fclearing$ on $\Sim$'s $(\Close,\cnt_\Dep,\cnt_\FWdr,S,b')$ passes:
          \begin{itemize}
            \item The bounds on the counts are $\Fvalidity$'s by \ref{inv:dw}, and $\Fclearing$ chooses the same requests, with the same $d_A$ and $w_A$.
            \item Claim~\ref{clm:sender}\ref{cl:debit} gives $S[A][B]\le\Auth_\cur[A][B]+\Carry_{\cur-1}[A][B]+\Pipe_\cur[A][B]$ for every honest $A$ and every $B$.
            \item $b[A]\ge0$ for every $A$, as $\Fvalidity$ computed $b$ from $\hlt{\Bal}=\Bal$, $S$, and the same deposits by \ref{inv:dw}, and required it. The new balance $b'[A]$ is at most $b[A]$, and the payout is $p[A]=b[A]-b'[A]=\min(b[A],w_A+W[A])\ge\min(w_A,b[A])$, as $W[A]\ge0$.
            \item For an honest $A$, $A$ signed every $q$ with $(q,\sigma)\in W$ naming $A$, by \ref{hyb:forge}. These $q$ are pairwise distinct requests for $\eid=\cur$, so $W[A]\le\Wdr_\cur[A]$ by \ref{inv:auth}, and $p[A]\le w_A+\Wdr_\cur[A]$.
          \end{itemize}
          So $\Fclearing$ pays out $\Fvalidity$'s $p$. Like the simulated ledger, it removes the same requests, sets $\Bal\gets b'$, and gives the same $\mathsf{Paid}$ and $\mathsf{Finalized}$ outputs. It closes $\eid$ with $S=\hlt{S_\eid}$, and sets $\cur\gets\eid+1=\hlt{\nxt}$ and its counters as the ledger does. This preserves \ref{inv:dw} and the epochs in \ref{inv:epoch}, and \ref{inv:auth} and \ref{inv:rcpt} as $\cur$ only grows.

    \item \emph{Honest senders check the close.} After the close is posted, each honest $A$ with a pending request $r$ does the following. If $r$ may settle in $\eid$, $A$ picks a receiver $B$ in $X_r$ and queries $\Fvalidity$ on $(\MTOpen,\eid,(A,B))$. If the opening shows the amount that the state of $r$ for $\eid$ gives $B$, $A$ books $r$ in $\eid$, that is, it marks $r$ as answered and sets $\cP_{A,\eid}$ to the map of that state. Otherwise $A$ drops $r$ if every epoch it may settle in is at most $\eid$. Claim~\ref{clm:sender} covers these steps, which produce no output.

    \item \emph{Honest receivers challenge.} $\Fclearing$ halts right after the close iff $S[A][B]<\Rcpt_\eid[A][B]$ for some $A$ and some honest $B$. When the close is posted, each honest $B$ queries $\MTOpen$ for every $A$ with $\hlt{\Rcpt^B_\eid}[A]>0$ and obtains an opening of $S[A][B]$, or a non-membership proof if $S[A][B]=0$. As \ref{inv:rcpt} held before the close, $B$ challenges iff $S[A][B]<\Rcpt_\eid[A][B]$. Its stored receipt verifies, so the simulated ledger halts on the spot. So an honest challenge halts the simulated ledger iff $\Fclearing$ halts after the close, which preserves the rest of \ref{inv:epoch}.
  \end{itemize}

  \paragraph{Receipts.} Let an honest $B$ accept a receipt from $A$ for epoch $\eid$ with amount $y$. If $A$ is corrupt too, $\Release$ imposes no bound. Otherwise Claim~\ref{clm:sender}\ref{cl:receipt} gives $y\le\Auth_\eid[A][B]+\Pipe_{\eid-1}[A][B]+\Pipe_\eid[A][B]$. So $\Fclearing$ accepts $\Sim$'s $\Release$, and the receipt step of both cases applies.

  \paragraph{Challenges by corrupt parties.} When a challenge by a corrupt party halts the simulated ledger, $\Sim$ sends $\Fault$ if $\Fclearing$ has not already halted. $\Fclearing$ accepts it, as $P_\op$ is corrupt, and both have halted. If $\Fclearing$ has already halted, it did so in the close just posted, and already gave the $\mathsf{Halted}$ and $\Refund$ outputs that the simulated ledger gives now.

  Hence $\REAL_{\Pi,\adv,\cZ}[1^\secpar,z,M]\approx_c\IDEAL_{\Fclearing,\Sim^\adv,\cZ}[1^\secpar,z,M]$ for every PPT $\cZ$, every $z$, and every $M$ not containing $P_L$.
\end{proof}